\documentclass[11pt]{article}
\usepackage{caption}
\usepackage{subcaption}
\usepackage{amsfonts}
\usepackage{bbm}
\usepackage{amsmath}
\usepackage[ruled,vlined,linesnumbered]{algorithm2e}
\usepackage[section]{placeins}
\usepackage{epsfig}
\usepackage{amssymb}
\usepackage[numbers]{natbib}
\usepackage{graphicx}
\usepackage{amsthm}

\usepackage{booktabs}
\usepackage{etoolbox}
\usepackage[round-mode=places,detect-weight=true,detect-inline-weight=math]{siunitx}

\theoremstyle{definition}
\newtheorem{definition}{Definition}[section]

\SetKwInput{KwInput}{Input}                
\SetKwInput{KwOutput}{Output}              

\begin{document}

\title{Anomaly Detection using Principles of Human Perception}

\author{Nassir Mohammad \\ \emph{Cyber Innovation and Scouting, Airbus, Newport, UK}}

\maketitle

\begin{abstract}
In the fields of statistics and unsupervised machine learning a fundamental and well-studied problem is anomaly detection.  Anomalies are difficult to define, yet many algorithms have been proposed. Underlying the approaches is the nebulous understanding that anomalies are rare, unusual or inconsistent with the majority of data. The present work provides a philosophical treatise to clearly define anomalies and develops an algorithm for their efficient detection with minimal user intervention. Inspired by the Gestalt School of Psychology and the Helmholtz principle of human perception, anomalies are assumed to be observations that are unexpected to occur with respect to certain groupings made by the majority of the data. Under appropriate random variable modelling anomalies are directly found in a set of data by a uniform and independent random assumption of the distribution of constituent elements of the observations, with anomalies corresponding to those observations where the expectation of the number of occurrences of the elements in a given view is $<1$. Starting from fundamental principles of human perception an unsupervised anomaly detection algorithm is developed that is simple, real-time and parameter-free. Experiments suggest it as a competing choice for univariate data with promising results on the detection of global anomalies in multivariate data. 
\end{abstract}

%
%
%
%
%


%
\section{Introduction} 
%

Anomaly---also referred to as outlier---detection is a well-studied subject where anomalies are commonly understood to be events or patterns that are unusual or unexpectedly different from normal instances. The importance of detecting anomalies is prevalent in wide ranging domains where any kind of information processing is required, and includes psychology, statistics, cyber-security, medicine, finance, agriculture and image processing to name a few. The identification and removal of anomalies can aid in understanding the data through visualisation and descriptive statistics such as the mean and standard deviation, where otherwise such anomalies can distort their meaning and interpretation. Machine learning, being increasingly utilised across many disciplines, also benefits from the removal of anomalies in order to better prediction and classification performance. Anomalies themselves are also of interest and may actually contain the most valuable information. For example, in the medical field once anomalies are detected during diagnosis they can be further analysed to gain theoretical and practical insight, while in the cyber-security domain the detection of anomalies can point to abnormal states, system intrusions or unusual characteristics that deserve further human investigation. 

Anomaly detection has its roots in the statistical community, but with advances in computing technology over recent decades it has become a field where computer scientists have increasingly participated. The latter have focused their efforts on attaining practical outputs as opposed to mathematical derivation and precision, and especially where large data is involved, computational efficiency, scalability, accuracy, interpretability and intuitive analysis have taken precedence. 
The problem usually falls in the realm of unsupervised learning where a learning algorithm only has access to data without classification labels. However, in practical scenarios we usually have access to a tiny amount of labelled data that can be intuitively helpful in building or validating a model although this is not enough to do supervised learning or to build a model of normality; thus algorithms assume no access to data labels. 

One of the main difficulties encountered in developing anomaly detection algorithms stems from defining what an anomaly is, this has remained elusive with the term subjective and imprecise. Indeed, practitioners often refer to the problem as detecting `unusual activities', or identifying `rare and inconsistent' observations or finding points that are `relatively distant' or `isolated' from a majority of points in multidimensional vectorial space. Most algorithms approach the problem by heuristic means, and model the normal instances first, before identifying everything outside as anomalous. This optimises the models for the normal data and can result in unsatisfactory performance for the real problem of detecting anomalies. A few algorithms however, \emph{directly target} the anomalies under the assumption that they are \emph{relatively few and sufficiently different} from the majority of normal instances. Although such intuitive understanding has led to the development of many successful algorithms, existing methods are not inspired by the perception of anomalies by the human visual system. While this is not a problem in and of itself, the human capability to detect anomalies is excellent and is arguably the most versatile and reliable; I believe it to be beneficial to at least be inspired by our knowledge of human perception. 

Another crucial problem of most, if not all, anomaly detection algorithms is that they are laden with one or more parameters that are left to the user to specify and adjust. In the case of supervised learning this may be reasonable and effective but can still be an expensive and slow process due to the large parameter space that may have to be searched. In the case of unsupervised learning where we have (little or) no access to labelled data the selection of appropriate parameters is even more challenging. The problem is exacerbated by the fact that the hyper-parameters are a fragile component that is sensitive and crucial to the models; performance on one data set may not generalise to another, and reproduction of results and comparison of algorithms is made difficult. For example, the contamination ratio parameter that many anomaly detection algorithms ask users to specify sets the percentage of data that is labelled by the algorithm as anomalous. This is difficult to determine in advance in unsupervised learning scenarios, and even if the correct value is chosen for training data, concept drift and the fact that anomalies are rare mean that new data sets are not guaranteed to contain the same percentage---indeed if any---of anomalies every time. Such problems indicate that the modelling has perhaps failed to capture some hidden principle and thus burdening users with the problem of parameter selection. Furthermore, where models require parameters to be obtained or guessed from the data using non-automated methods or human intuition, it could be argued that they fall out of the \emph{truly} unsupervised realm of machine learning and into some level of supervision, even if the data labels are still unknown to the algorithm.

The first contribution of the present work is to address the problem of defining anomalies and to develop a philosophical approach for their detection that is inspired by the human visual system. Specifically, the connections between the Gestalt Theory of human perception and the identification of unexpected events will be explored by starting from some good principles, primarily the Helmholtz principle formally stated by \citet{Lowe85}: \emph{every structure that shows too much deviation from random noise calls attention and becomes a perception}. The central idea is that data be represented appropriately by features and then into numerical values, where each value represents both an individual window and the number of \emph{indicators} that are viewed within that window, such that given an assumed uniform distribution of all the indicators across all windows, every unexpectedly large value corresponds to anomalies of interest. The present aim is to detect \emph{global} anomalies as opposed to local anomalies, where the former refers to observations falling in the outer regions of a single group and the latter refers to anomalies that may be buried in between distinct groups or inside a group of normal observations---their detection is left to future work. (However, some of the illustrations will show local anomalies when the data is considered as a whole, this is for explanatory purposes.) Thus, given a grouping of observations by say relative proximity, any observation that lies sufficiently far from this main grouping is at once perceived to be something unusual, unexpected, with respect to the grouping. This approach yields the second contribution of the paper in presenting the \emph{perception} algorithm; a \emph{truly} unsupervised parameter-free algorithm for all gestalt grouping laws that enables \emph{direct} computing of anomalies with respect to the different representations of data. We are not necessarily restricted to just visual problems and analysis, and indeed will generalise the approach to derive \emph{meaningful events} (whose expectation of occurrence is $<1$) that correspond to the detection of anomalies in the general case. 

I take a path that marries the statistical and the computational approaches together. The former can explain why a particular approach is working by starting from good principles and simple assumptions to derive mathematical representations and an algorithm for solving the tasks, while the latter guides to develop solutions that are feasible and applicable in modern production environments. The requirement is to have a mathematically derived and well understood real-time, memory efficient, robust, reliable and parameter-free algorithm that is not restricted by data size or dimensions, with performance that matches or betters current methods. The results in section \ref{section:implementation} show that the perception algorithm performs remarkably well against existing univariate anomaly or outlier detection algorithms that are used across the sciences. Encouraging conclusions are also reached for the first multivariate version of the algorithm provided some data assumptions are satisfied. 

To facilitate the exposition, the rest of the paper is laid out as follows: Section \ref{section:review} reviews some of the best known and widely used anomaly detection algorithms. Due to the complexity of reviewing and evaluating algorithms, the descriptions are kept to a few key properties such as computational efficiency, parameters and ease of use. Section \ref{philosophy} presents the philosophy of the anomaly detection approach that is inspired by the Gestalt Theory of human perception. Section \ref{section:motivation} provides a couple of motivating examples for algorithm development. The first is the classical \emph{birthdays in a class problem} where it is explained why expectations of random variables are preferred over probabilities. Second is an example on \emph{tossing coin sequences} where exceptionality of certain sequences that we perceive---corresponding to gestalt groupings---of low probability are calculated. This example shows that we are primarily concerned about certain a priori groupings, and how to handle the probability thresholding problem using expectations. Then the Helmholtz principle is stated in a generic way and $\varepsilon$-meaningful events are defined for the general detection of exceptional events. Section \ref{section:algorithm_description} discusses a real cyber-security intrusion detection scenario that inspired and helped develop the perception algorithm which is detailed. This is a generalisation of the solutions to some of the questions in the motivating examples, and a simple extension for handling multivariate data is provided. Section \ref{section:implementation} gives the implementation details, results and evaluation for wide ranging univariate and multivariate data. Finally section \ref{section:conclusions} concludes the paper together with thoughts for future work.


\section{Review of Anomaly Detection Methods}
\label{section:review}

Anomaly detection algorithms each have their relative strengths and weaknesses and application domains. The literature is vast and others have provided excellent and comprehensive reviews: \cite{Chandola2009}, \cite{Goldstein2016}, \citep{AggarwalOutlierAnalysis2017}. Thus, in this section only properties of some well known and widely adopted algorithms are reviewed rather than providing a thorough analysis. However, this concise review also provides a valuable reference to methods that have crept to the top of literature reviews, articles and real-world application. Note that only relatively simple and perhaps the first algorithms for a particular technique are discussed since it is these that have been often found to perform very well. More complex variations and algorithms do not always yield the performance gains that one would expect from their sophistication \citep{AggarwalOutlierAnalysis2017, AggarawalEnsembles2017}. The techniques that carry out anomaly detection include statistical (parametric and non-parametric), distance, density, clustering, classification and isolation based methods. An important commonality of the algorithms that deserves special mention is that users are asked, even in the simplest of problems, to specify parameters that are crucial to the performance of the models. This is difficult to do in practice due to the unsupervised nature of the problems that are of concern. Other important considerations are assumptions of the data distribution, the computational complexity of the algorithms for iterative data processing, feasibility of application, and also the ease of use. 
 
\subsection{Statistical Techniques}
\label{section:statistical_techniques}
Statistical methods are some of the simplest and most widely used for univariate anomaly detection. Methods include z-scores, modified z-scores, IQR (boxplots), Grubbs test, Dixon’s test and the Shapiro-Wilk W test \citep{iglewicz1993detect}. Statistical methods are derived mathematically, with parametric methods making assumptions about the data distribution while non-parametric methods make none. They are generally easy to implement, fast to train and test with once models are built. However, parametric models can enter difficulties if the data distribution does not match the assumption---which is usually the Gaussian distribution. Non-robust measures also have high sensitivity to outliers (like the commonly used z-score method) that can lead to increased false-negatives. The methods are not directly suited to multivariate data which hinders their application for many machine learning problems. They also require users to specify parameters such as the number of standard deviations from the mean, number of median absolute deviations, the number of outliers in the data and/or their positions, or multiplications of interquartile ranges in order to specify thresholds. These are normally selected as a `rule of thumb' or based on judgement and experience but can and do vary in the literature, with the results being significantly different depending on the chosen parameters. 

Another statistically related method is the histogram based outlier score (HBOS) \citep{goldstein2012} which is a relatively recent, simple and efficient unsupervised learning algorithm that only assumes independence of the features. It is directly applicable to multivariate data and calculates the degree to which an observation is anomalous by building histograms for each feature, and then for each observation the inverse height of the bins it resides in (representing the density estimation) of all features are multiplied. The idea is very similar to the Naive Bayes algorithm in classification where all independent feature probabilities are multiplied. The main advantage of the algorithm is its (linear with data size) speed and accuracy on detecting \emph{global} anomalies that rivals more complex methods. However, it's neglecting of dependencies between features does mean that it may not capture anomalies with respect to the data axes and hence feature engineering could be required to expose anomalies; although this becomes less severe with higher dimensional data sets due to larger sparsity. The algorithm also requires a few parameters to be specified in advance by users including the number of bins $b$ to be used and the ratio of anomalies assumed in the data set.

\subsection{Distance and Density Based Techniques}
\label{section:distance_and_density_techniques}
Nearest neighbour based algorithms assume that normal data instances occur in dense neighbourhoods and anomalies occur in sparse regions or far away from closest neighbours. The basic distance based unsupervised $k^{th}$ nearest neighbours ($kNN$) algorithm \citep{Ramaswamy2000} is straightforward and effective for detecting anomalies and does not make any assumptions about the generative distribution. Of the density based methods local outlier factor (LOF) \citep{Breunig2000} is the most well known non-parametric algorithm where normal data form dense regions and anomalous data fall in sparse regions. LOF is adept at detecting local outliers in point anomaly detection, and is among the best performing algorithms for multivariate data. However, for both distance and density based methods it is crucial to choose an appropriate parameter $k$---for the $k$ nearest neighbours---in order for the algorithms to perform well and also the correct ratio of anomalies assumed to be in the data set. Generally it is challenging to choose the appropriate parameters, especially in unsupervised settings. The standard techniques are also computationally expensive, difficult to implement for data streams and challenging in high dimensional data. The requirement of finding the nearest neighbour, which all these methods rely on, results in $O(n^2)$ complexity with the number of examples in the data $n$. This can make iterative development tedious and the algorithms unable to complete within reasonable time on larger data sets.
 
\subsection{Cluster Based Techniques}
\label{section:cluster_techniques}
The underlying assumptions in clustering algorithms include the following: that normal data belongs to clusters while anomalies do not belong to any; or anomalies are far away from cluster centroids; or anomalies belong to small or sparse clusters. The general method is to cluster the data and calculate the centroids or density of each one found. A data point can then be classified (possibly by another algorithm) either according to the distance it is from a relatively known large cluster (too far and it is an anomaly while close enough it is classed as normal), or based on the relative density around the point and it’s neighbours. A goto method is the $k$-means algorithm \citep{macqueen1967} due its simplicity and efficiency, and DBSCAN \citep{Ester96} is also a popular but slightly more complex method that has demonstrated practical real-world use for clustering. Some key advantages of both (usually via some enhancements) are their non-parametric nature, detection of both local and global anomalies, ability to be used in incremental mode for data streams, and being typically faster than $O(n^2)$ with data size $n$. The main disadvantages of both algorithms is that anomalies are detected as by-products of clustering and thus their detection is not optimised. Indeed practical results can be somewhat unsatisfactory. The user is also left with the challenging task of possibly choosing several algorithm parameters that are important for success. The $k$-means algorithm also assumes clusters are spherically distributed and being non-deterministic may have to be run several times to achieve stable results which can negate the algorithm speed and ease of use. On the other hand, DBSCAN, while having the advantage of being able to find arbitrarily shaped clusters, can struggle with large differences in densities.    
 
\subsection{Classification Techniques}
\label{section:semi_supervised_techniques}
The one class support vector machine (OCSVM) \citep{sch99} can be a powerful unsupervised anomaly detection algorithm that fits a model to only one class. Like the traditional supervised SVM the model creates a decision boundary (a hypersphere) that can be used to classify future incoming data points. It is more suited for novelty detection where it is assumed that there are no anomalies in the training set or that the training set has been thoroughly cleaned to minimise their affects. A soft margin can also be used to make it more robust to outliers. The algorithm excels in multimodal, non-Gaussian distributed and high dimensional data sets. However, the OCSVM requires a number of parameters to be set by users such as the type of kernel (e.g. linear or rbf), $\nu$ that approximates the ratio of anomalies to be handled by a soft-margin, $\gamma$ the kernel coefficient, and the ratio of anomalies to be returned from the data set when using outlying scores; these parameters are data dependent and important for optimal results. The complexity is difficult to determine since it depends on the chosen parameters and the number of support vectors. In particular, when training sizes are large in the number of examples, kernels become computationally burdensome to use and a linear decision boundary may be learnt instead. Another important point to keep in consideration is that outliers that slip past the model’s detection can cause gradual degradation of accuracy.  

Another powerful and elegant way to detect anomalies is to assume the majority of the data follows a Gaussian distribution and apply a multivariate Gaussian distribution (MVG) fit using robust (i.e. with mitigation of outlier influence) estimates of the covariance matrix; thus fitting an ellipse to the central data points while ignoring points further outside. This method is also known as the elliptic envelope \citep{scikitlearn} calculated using the minimum covariance determinant (MCD) algorithm \citep{mcd99}. Anomalies are detected as points that lie in low probability regions of the space by using Mahalanobis distances obtained from the covariance estimate. An advantage of this method is that it automatically captures the correlation between features, however this comes at the cost of being relatively more computationally expensive when compared to simple statistical models or where independence of features is assumed. The algorithm does not scale well with the number of examples $n$, which also must be greater than the number of features $d$, otherwise the covariance matrix is not invertible and the calculation will fail. Due to such problems in higher dimensional spaces, it is advised to ensure $n > d^{2}$. The MVG model also requires users to specify in advance the ratio of anomalies expected to be in the data set. This is an important parameter as it informs the classifier of the proportion of outliers to be returned and has to be guessed at using expert domain knowledge or exploration of the data.  

\subsection{Isolation Techniques}
\label{section:tree_techniques}
Random forests and their variations have a reputation for being amongst the best machine learning algorithms for structured data. They work well with different data repartitions and are efficient with high dimensional data due to the fact that they are algorithm trees, where the structure provides more efficiency compared to models that use distance based computations. Furthermore, at classification time the number of feature comparisons to make is equal to the height of the tree making them highly applicable for real-time anomaly detection. The isolation forest algorithm \citep{Liu2008} takes advantage of this tree like structure to explicitly isolate anomalies instead of profiling normal instances. The algorithm trains many partial models on random subsamples of data by iterating through the training set, randomly selecting a feature and randomly selecting a split value between the maximum and minimum values of that feature. The intuition behind this method is that inliers have more feature value similarities which require them to go through more splits to be isolated, while outliers are easier to isolate with a smaller number of splits because they will likely have some feature value differences that distinguish them from inliers. By measuring the path length from the root of the tree every point is given an anomaly score where the shorter the path length the more anomalous the data point. The algorithm claims a linear time complexity with a low constant and a low memory requirement and produces amongst the best results in practice. Being fast to run and scaling well with the number of features and data size, it is easy to implement and use for iterative development. Drawbacks of the technique however are that it performs only axis aligned cuts of the data which might lead to suboptimal results, and users must specify in advance the ratio of anomalies expected to be in the data set; however, the remaining set of the parameters can be left at default values to give optimal results.

%
\section{Philosophy of the Approach}  
\label{philosophy}

There are a number of questions that one likely asks in their initial study of anomaly detection, particularly from the human visual perspective which is often how we describe and understand anomalies. Here are a few of importance and relevance:

\begin{enumerate}
\itshape{ 
\item What exactly is an anomaly, and hence anomaly detection?
\item How are anomalies perceived by the visual system?
\item Why do we interpret some collections or points---stimuli that arrive at our eyes---as groups and anomalies and why is it important at a primitive level?
}
\end{enumerate}

In order to start addressing the first question, consider the following prevalent definitions of the two most commonly used terms of \emph{outliers} and \emph{anomalies}:
\begin{itemize}
\itshape
\item An outlier is an observation which deviates so much from the other observations as to arouse suspicions that it was generated by a different mechanism \citep{Hawkins80}.  
\item Anomalies are instances or collections of data that occur very rarely in the data set and whose features differ significantly from most of the data. 
\item An outlier is an observation (or subset of observations) which appears to be inconsistent with the remainder of that set of data \citep{Barnett1978}.
\item An anomaly is a point or collection of points that is relatively distant from other points in multi-dimensional space of features.
\item Anomalies are patterns in data that do not conform to a well defined notion of normal behaviour \citep{Chandola2009}. 
\item Let $T$ be observations from a univariate Gaussian distribution and $O$ a point from $T$. Then the z-score for $O$ is greater than a pre-selected threshold iff $O$ is an outlier.

\end{itemize}
The definitions are either imprecise---they don't tell us exactly \emph{what} to compute---or specific to a method. Difficulties that arise here include how do we formalise normal behaviour, or define inconsistency, or give a measure of relative distance or select a statistical threshold. Furthermore, it is clear there is no universally accepted way of defining a single or collection of observations as an anomaly, rather there can be many explanations---singular or plural---as to why we can class an observation as anomalous. 

Before continuing, let us differentiate between the commonly understood notions of \emph{outliers} and \emph{anomalies} to be used in this paper. Intuitively outliers are taken to be those observations that are inconsistent, unusual or unexpectedly different from some other group of observations (to be defined formally later) while anomalies are \emph{those outliers} that are of interest to the task or observer. This subtle difference between them is important since many outlier observations can be extracted from data, yet be of no semantic interest even if they satisfy the requirements of being relatively rare and differing significantly from most of the data. As a crude example, consider a building that needs to be monitored by a security team who are tasked with simply reporting any outliers. To this end they could perceive many outliers to the minutest details in the building---from carpet tears, to unusual paintwork or a cold spot in a room---but which would be of no interest from a security perspective. Thus the outliers of interest have to be detected and chosen with respect to features of interest such as functioning door locks, suspicious movements of personnel or events at particular times of the day. This paper is concerned with detecting anomalies, but general outlier detection techniques will also be discussed, particularly when referencing other works, since the semantic information may not always be apparent and the words are often used interchangeably in the literature. Naturally, also, if we have a univariate data set, then we must already assume that we are working with the feature of interest and thus outlier and anomaly detection would coincide. 

\begin{figure}
\centering
  \includegraphics[width=0.6\linewidth]{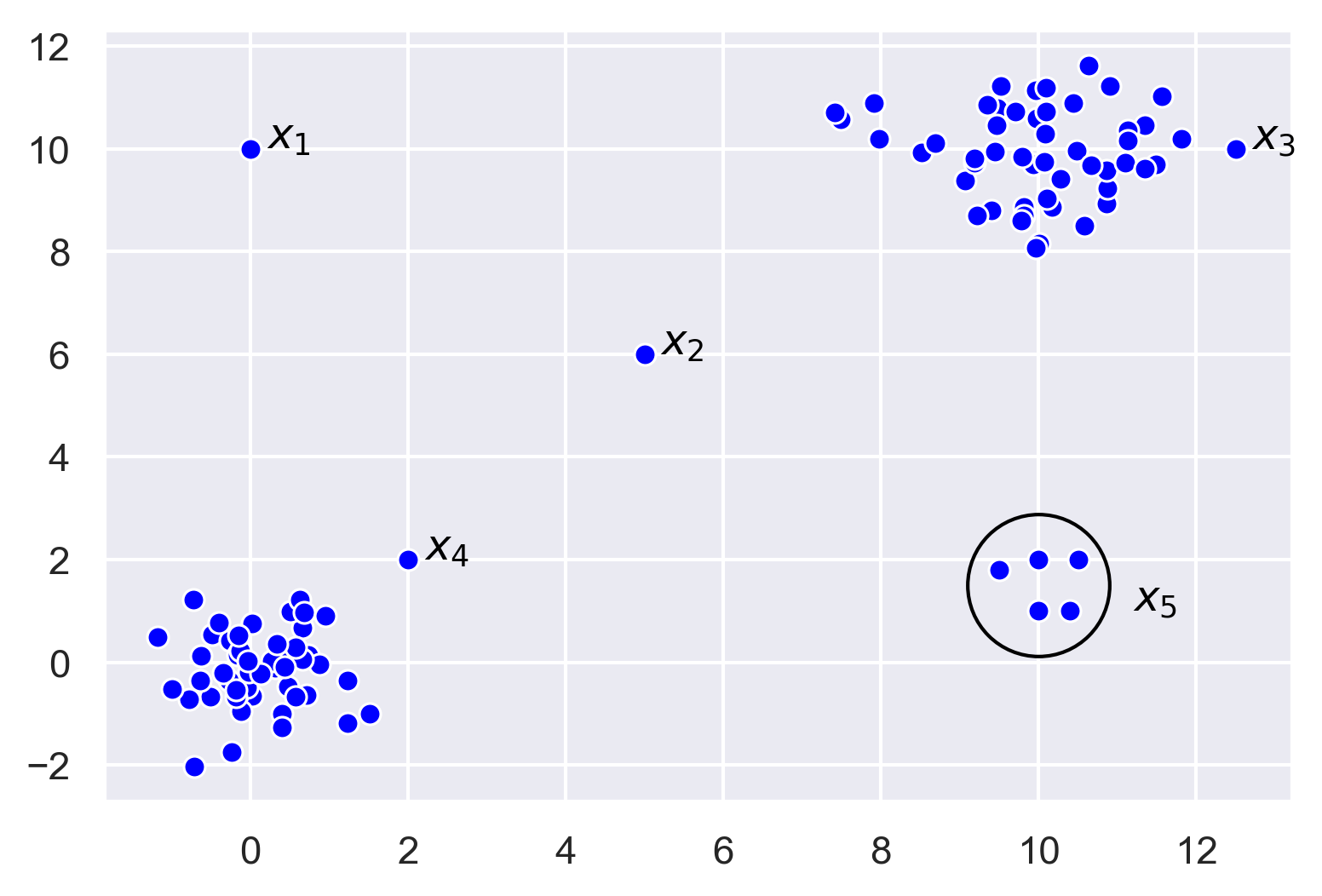}
  \caption{Subjectivity of detecting anomalies; which points do we consider to be an anomaly? Certainly $x_1$ and $x_2$, but what about points $x_3$, $x_4$ and the small cluster of points $x_5$? How do we define the threshold for a point to be an anomaly, and why?}
  \label{fig:anomaly_subjectivity}
\end{figure}

One of the aims in this paper is to provide a precise definition of what constitutes an anomaly---hence answering the first question of this section. However, in order to do so aspects of the second question on how the visual system perceives unusual observations must be dealt with; this is challenging to answer, if it can be answered at all under our current understanding. Indeed, one could say that \emph{how} groups of pixels (retina cells in the eye or pixels in a camera) form into spatial objects and then meaningful entities is one of the major enigmas of perception. The detection of anomalies is assuredly subjective and many explanations can be given for why an observation could be considered an anomaly, and even why the same observation could be considered to be perfectly normal (usually driven by the context or past experience of the observer but not necessarily). This is exemplified by Figure \ref{fig:anomaly_subjectivity} where we can ask \emph{which points an observer would label an anomaly?} Depending on the observer's experience or interpretation, different labels could be given to different points. Although we would be sure that $x_1$ and $x_2$ should be classed as anomalies; $x_3$, $x_4$ and the small cluster $x_5$ are more subjective and perhaps require domain knowledge to arrive at a decision. \emph{How do we define the threshold to class a point as an anomaly, and why?} As it is for our general understanding of vision, we don’t have definite answers for \emph{how} and \emph{why} we perceive anomalies. The principles of visual reconstruction are still beyond us, but there have been attempts to state its laws and to better understand the visual information processing behaviour. 

One scientific attempt to understand the visual reconstruction process and how to arrive at global percepts is the Gestalt Theory of Psychology based on the work of Max Wertheimer, Wolfgang Köhler, and Kurt Koffka. The gestalt psychologists believed that the sum is not only more---or different---than the parts, but that phenomena is organised as structured wholes which have priority over the parts. The theories of perception are based on human nature being inclined to understand objects as an entire structure rather than the sum of its parts. The whole defines the structure of the parts, rather than vice-versa. For example, when seeing a picture of a cat, the entirety of the animal is perceived immediately to \emph{see} a cat where apprehension is not through a process of part summation. Indeed, only after perception of the whole, the various parts of the animal may be considered, such as its whiskers, eyes or paws. In the fundamental work by \citet*{Morel07}, they describe the Gestalt program of which I provide an excerpt:

\textit{“The program of this school is first given in Max Wertheimer’s \cite{Wer23} founding paper. In the Wertheimer program there are two kinds of organising laws. The first kind are grouping laws, which, starting from the atomic local level, recursively construct larger groups in the perceived image. Each grouping law focuses on a single quality (color, shape, direction...). The second kind are principles governing the collaboration and conflicts of gestalt laws. In its 1975 last edition, the gestalt `Bible' Gesetze des Sehens, Wolfgang Metzger \cite{Metz75} gave a broad overview of the results of 50 years of research. It yielded an extensive classification of grouping laws and many insights about more general gestalt principles governing the interaction (collaboration and conflicts) of grouping laws. These results rely on an incredibly rich and imaginative collection of test figures demonstrating those laws...Computer Vision did not at first use the Gestalt Theory results: David Marr's \cite{Mar82} founding book involves much more neurophysiology than phenomenology...” }

\citet*{Morel07} tentatively translate the Wertheimer program into a mathematics and computer vision program. Addressing two fundamental matters of image sampling and information measurements, they translate qualitative phenomenological observations into quantitative laws with numerical simulations. They take a few basic principles---Shannon Sampling, Wertheimer's Contrast Invariance, Isotropy and the Helmholtz principle---to yield image processing algorithms that can be run on any digital image parameter-free i.e. not dependent on human intervention to provide critical parameters or tuning. These algorithms are developed to detect \emph{partial} gestalts which are the sub-groupings (elementary perception building blocks) that come together to form the \emph{global} gestalts (complete object perceptions). Focusing on the former, they develop a numerical framework for detecting edges, alignments of image segments and object borders that correspond to these \emph{partial} gestalts in digital images. This approach to the computer vision task of edge detection could be said to have parallels with the problem of anomaly detection. Here, detecting a large enough unexpected grouping of pixels in an image to form relatively rare edges is a difficult task, and the exact definition of an edge is subjective, while noise also acts as a hindrance to their computational detection---since both edges and noise occur where there are large variations in image intensity. 

Desolneux, Moisan and Morel are guided by the fact that a few good principles lead to quantitative laws in mechanics with exact predictions based on formal or numerical calculations. Most of the mathematical work is in formalising the Helmholtz principle. This attempts to describe when perception decides to group objects according to some quality, and can be understood in a number of subtle ways: \emph{gestalts are sets of points whose (geometric regular) spatial arrangement could not occur in noise \citep{Lowe85}; there is no perception in white noise; we immediately perceive whatever could not happen by chance; every structure that shows too much geometric regularity to be found by chance in noise calls attention and becomes a perception; whenever some large deviation from randomness occurs, a structure is perceived \citep*{Morel07}}.

\begin{figure}
\centering
  \includegraphics[width=1\linewidth]{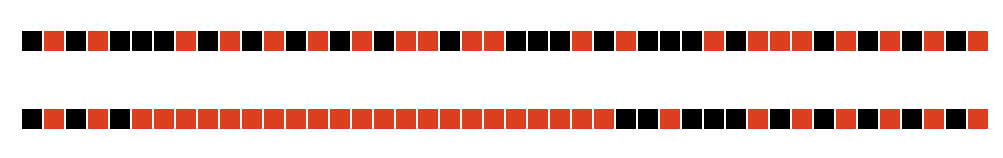}
  \caption{Two sequences of red and black squares are shown where each colour can occur with equal probability. The top sequence shows a realisation of possible outcomes where nothing notable is perceived. In the bottom sequence there are $22$ consecutive red squares which are immediately perceived as something that is unlikely to have occurred by chance given the total number of squares in a sequence. Here the Helmholtz principle and the colour constancy grouping law is at play so that a structure is perceived that is unlikely to happen in uniform random noise.} 
  \label{fig:red_black_squares}
\end{figure}

For an intuitive understanding consider Figure \ref{fig:red_black_squares} where a person observes a random sequence of red and black squares that can occur with equal probability. In the top row of the figure we observe a typically expected sequence, given the random nature for the occurrences of the squares. Here we do not perceive anything unusual. However, in the bottom sequence we have $22$ consecutive \emph{red} squares that is considered immediately noticeable. Here, the Helmholtz principle is at play where we perceive some structure that has very low probability to occur in uniform random noise. However, both sequences, and indeed all other possible red and black sequences of the same length have the same probability of occurrence, yet most of them appear without raising interest; only those corresponding to a \emph{grouping law}---here the colour constancy---calls attention. Thus, it is not enough to say that the sequence in question has low probability to occur, but we must also qualify our assessment with a \emph{relatively few} groupings (gestalts) of prior interest. The discovery and categorising of these grouping laws has been one of the main goals of the gestalt psychologists. The limitation on the number of perceptual grouping laws (perhaps by our interaction with the physical world or the statistical properties of images) is of paramount importance, otherwise we could not hope to perceive the relevant configurations from the enormously possible number of configurations of visual stimuli. 

With a preliminary understanding of the Helmholtz principle, the gestalt groupings and the difficulties in defining anomalies, I can proceed in answering the first and second questions at the beginning of this section. David Marr’s \cite{Mar82} book on vision is useful in this context. He points out in his introduction that to understand a complex device which performs an information processing task, one needs \emph{many different kinds of explanations}, and that for vision there is no single equation or view that explains everything. Each problem has to be addressed from several points of view. He states that to fully understand an information processing device one needs to understand it at different levels. The highest level is to understand \emph{precisely what} is being computed and \emph{why}: the computational theory. The second level concerns \emph{how} to carry out the computation and includes the choice of \emph{representation} for the input and output of the process, which has important consequences on what can easily be carried out and made explicit, and what is pushed into the background and maybe hard to recover. This level also contains the \emph{algorithm} by which inputs are transformed into outputs. There is usually a wide choice of representations, and the choice of representation often critically determines the choice of algorithm. Which one is chosen will usually depend on particular desirable or undesirable characteristics that the algorithm might have. The final and third level is the implementation mechanism describing how the representation and algorithm is physically realised. 

Marr uses a cash register to exemplify the levels of analysis. At the top level is precisely what a cash register does; it carries out arithmetic. The reasons for doing so are to combine prices of items to arrive at a final bill; \emph{constrained} by certain natural requirements of the task such as commutativity and associativity. The second level includes the choice of representation for the input and output of the problem. This is usually taken to be the Arabic numbers as opposed to say, Roman numerals or binary numbers, due to the implications it has on the algorithms that can be used to accomplish the calculating transformation from input to output. Finally, the third level is the physical realisation of the cash register used to implement the representation and algorithm. However, it could just as well be implemented by the human brain as a very different realisation. If we were to study the cash register from only a physical mechanical perspective, it would be difficult to understand the algorithm, but knowing about what it is computing and why, and about the nature of the problem it is trying to solve, we can have a much better and complete understanding of the device.

\begin{figure}
\centering
\begin{subfigure}{0.5 \textwidth}
  \centering
  \includegraphics[width=1\linewidth]{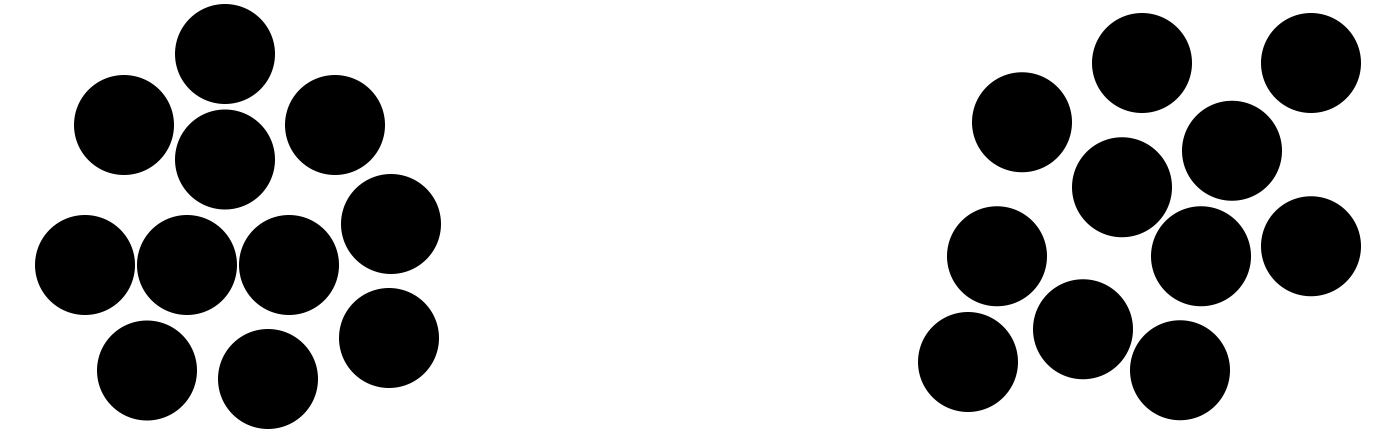}
  \caption{}
  \label{fig:vicinity_ex1}
\end{subfigure}
\par \bigskip 
\par \bigskip 
\par \bigskip 
\begin{subfigure}{1\textwidth}
  \centering
  \includegraphics[width=0.5\linewidth]{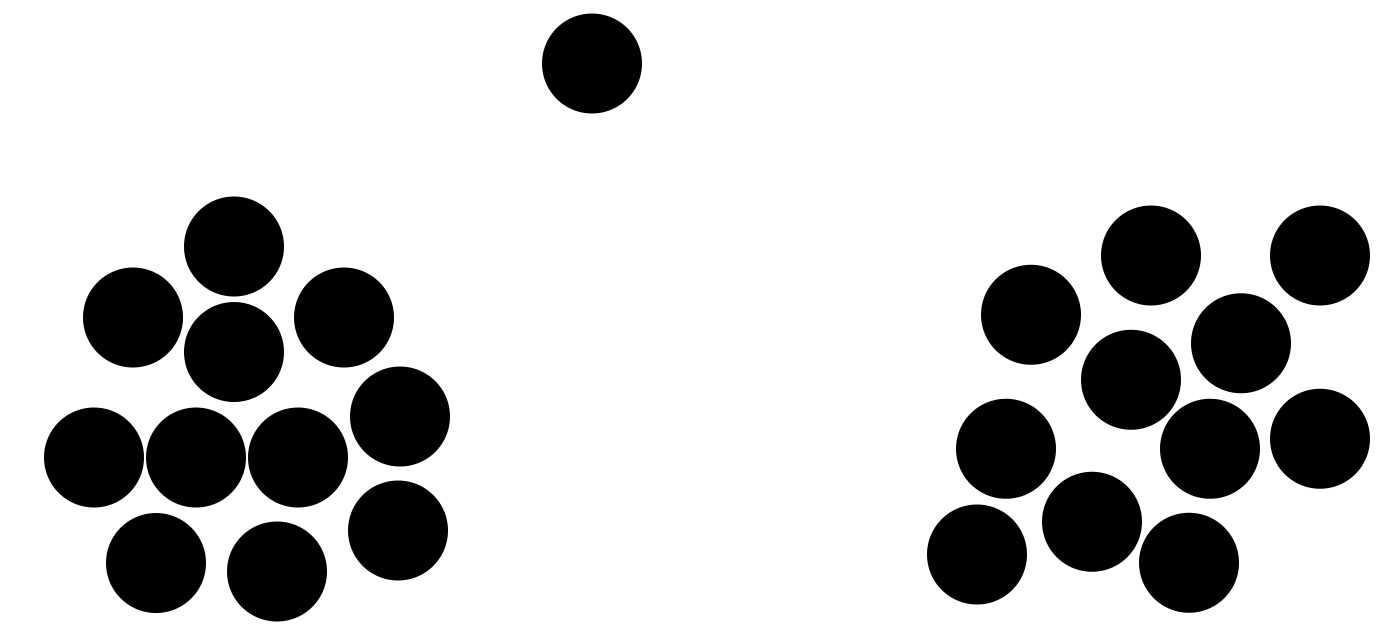}
  \caption{}
  \label{fig:vicinity_ex2}
\end{subfigure}
\caption{Vicinity (or proximity) law. (\subref{fig:vicinity_ex1}) Two groups of circles are perceived as two separate objects due to the distance between circles within clusters being similar, and much smaller than that between the two clusters. (\subref{fig:vicinity_ex2}) The same two clusters of circles but now with an additional observation. This is perceived to be an anomaly with respect to both groupings. 
}
\label{fig:vicinity_ex}
\end{figure}

\begin{figure}
\centering
\begin{subfigure}{1\textwidth}
  \centering
  \includegraphics[width=1\linewidth]{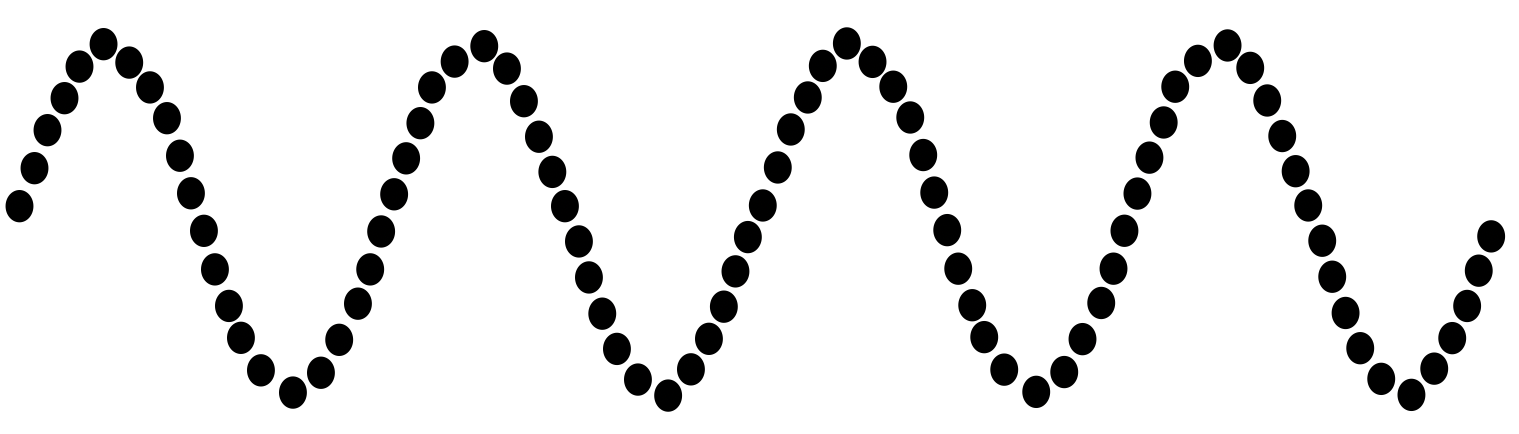}
  \caption{}
  \label{fig:good_continuity_ex1}
\end{subfigure}
\par \bigskip 
\begin{subfigure}{1\textwidth}
  \centering
  \includegraphics[width=1\linewidth]{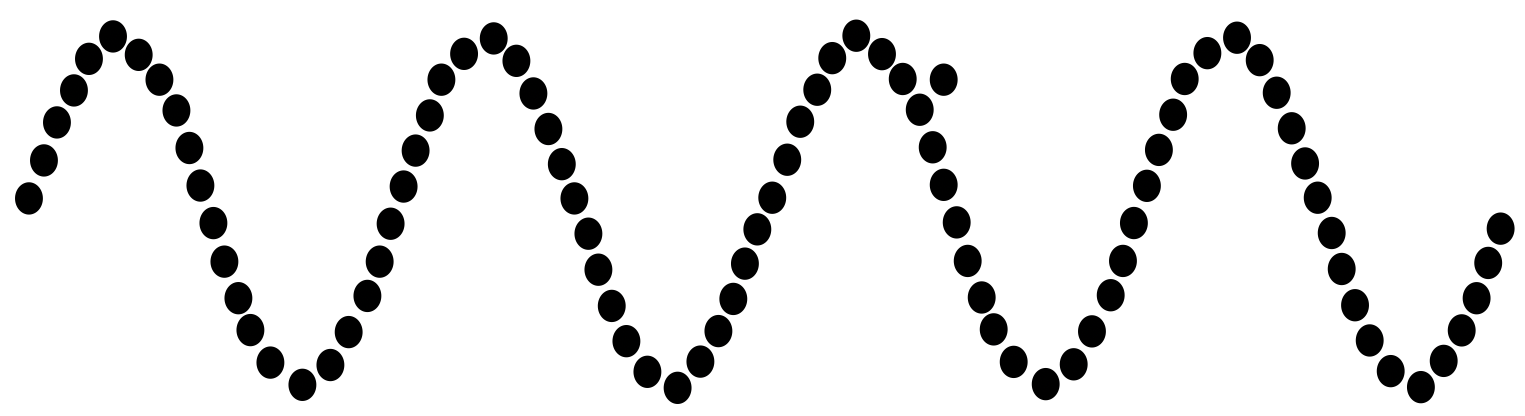}
  \caption{}
  \label{fig:good_continuity_ex2}
\end{subfigure}
\caption{Good continuity law. (\subref{fig:good_continuity_ex1}) A set of observations is shown that could be grouped by similarity of shape, colour constancy, or vicinity. However, the most apparent gestalt law in action is good continuity of the repeating pattern. (\subref{fig:good_continuity_ex2}) The same observations but with an additional circle which is not an anomaly with respect to similarity of shape, colour or local vicinity, but with respect to the good continuation of the repeating pattern.
}
\label{fig:good_continuity_ex}
\end{figure}

Marr's different perspectives of explanations and levels of analysis led to consider anomaly detection as an information processing task, and to propose that the anomaly detection problem is not necessarily ill-defined or entirely subjective, but rather there are many explanations as to what could lead to an observation or collection being classed as anomalous. I propose these explanations are actually limited in number and correspond to the gestalt groupings (although some transformation of the unperceived representation to one that is perceived may be required). This helps explain why the definition of an anomaly is so subjective and has been difficult to provide. There are many explanations for why humans perceive an observation as being an anomaly but the laws governing these explanations can be quantified into a relatively few groups, and perhaps fewer will be of central importance depending on what is being computed (often dictated to by the nature of the data and task). An object may be perceived to be an anomaly due to its colour being different, or its size being relatively smaller, or for being a relatively large distance from other points, or for not following a plane of good continuation, or for breaking the symmetry of the distribution of points, or for breaking an alternating pattern. The gestalt groupings form the representation of the observations, and given a single representation, the choice of algorithm to detect deviations from the grouping to be classed as anomalies can be many, e.g. based on distance, density or statistical distribution assumptions. An algorithm based on anomalies being separated by relatively large distances only may work well where the grouping of the observations satisfy these assumptions (see Figure \ref{fig:vicinity_ex}), but completely fail where observations are following a \emph{good continuity} grouping law (see Figure \ref{fig:good_continuity_ex}) and hence anomalies are not necessarily detectable just by their relative distance to other points, but instead by how they deviate from the good continuity. 

The Helmholtz principle can be used to explicitly derive formulas that find perceptual groupings, and anomalies can also then be uncovered by applying the same principle but with a complementary measure with respect to the grouping. In fact, anomalies can be detected directly without modelling the normal data and a first example of this is detailed by the perception algorithm in section \ref{section:algorithm_description}. The algorithm removes the subjective element of deciding if an observation is an anomaly for a given representation and measure through the use of statistical expected values. The present work is primarily concerned about anomalies with respect to a single grouping, however, the approach is equally applicable to deciding whether an observation is an anomaly with respect to local groupings that may come together to also form larger groupings in a hierarchal fashion. Furthermore, although the gestalt laws are usually associated with visual problems, the concept of groupings will be used in higher multidimensional space to derive anomaly detection algorithms for multivariate data.

The gestalt laws---for example similarity, vicinity, symmetry, colour, constancy and good continuity---and the Helmholtz principle help us understand \emph{how} we see anomalies. This also leads us to our definition of \emph{what} an anomaly is:

\begin{definition}[Anomaly with respect to the gestalt laws]
A grouping of interest represented by a gestalt law is perceived by the Helmholtz principle when it is unexpected to happen (i.e. its expectation of occurrence is $<1$) in uniform random noise. Any observation that is unexpected to occur with respect to this grouping is perceived, by the same principle, to be an anomaly.
\end{definition} 

Thus, we precisely see what is to be computed to detect anomalies, and yet the definition is left open enough to allow the construction of measures that can deem an observation to be anomalous with respect to the grouping. The reasons to propose this as a valid definition of what an anomaly is, and what is to be computed for anomaly detection is due to the assumption that humans decide innately or by bringing many features of interest from past experience whether an observation is anomalous by considering that which differs unexpectedly from the main grouping to be an anomaly. The natural reason for taking expectations $<1$ is that if the event is not expected to occur even once, but has occurred, then it is deemed unexpected; this will be further expanded upon in section \ref{section:motivation}. There is an analogy here with inductive reasoning where an observation that is unexpectedly deviant---an anomaly---can be considered as that which shatters the understanding formed from the majority normal observations. It is also assumed that anomalies are \emph{rare} and \emph{sufficiently different} to the normal data. These aspects are left imprecise but will not lead to difficulties since the conditions are naturally satisfied for the problem of anomaly detection, otherwise it would be considered to be a problem of classification. Such natural constraints ensure the problem is tractable.

The third question asked at the beginning of this section concerning why the human visual system perceives groupings and anomalies and what is its importance cannot be addressed satisfactorily yet. Presently, I can only summarise conjectures that the reasons why the list of gestalt groupings of geometric qualities is perceived may be directly related to the geometric formation and hence statistics of the natural world. Indeed, the statistics of natural images have been proposed to understand the visual systems of humans and animals by trying to understand the stimuli that such systems are optimised to process. By understanding the information processing task and what visual problems the brain is solving we can better understand its functioning. This approach has led to findings that support the processes in early vision and applications in computer vision. Regarding the importance of the detection of anomalies in vision, this may be attributed to unexpected events being those where most information and meaning is contained, and their identification is much more important than that which is expected to occur for our interaction with the world around us. 

The remainder of this section provides a handful of visual examples to clarify the overall approach and show the gestalt groupings at work; each one also illustrates anomalies at play. Whenever there are observations---elementary or complex---that share a similar characteristic, the human visual system naturally groups the observations to form a larger gestalt. Note that indeed, several gestalts may be at work in a single example, but we are interested in the dominating one. 

\begin{figure}
\centering
\begin{subfigure}{1\textwidth}
  \centering
  \includegraphics[width=1\linewidth]{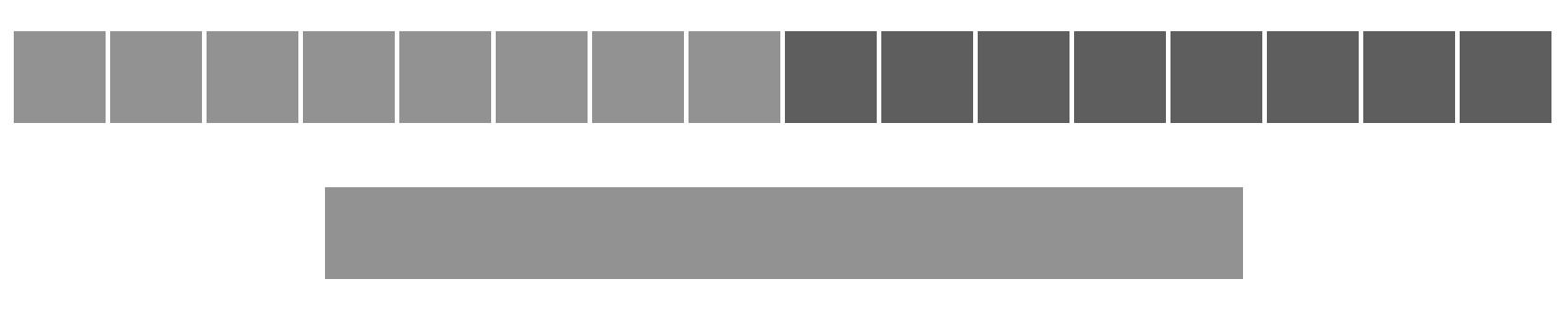}
  \caption{}
  \label{fig:colour_constancy_ex1}
\end{subfigure}
\par \bigskip 
\par \bigskip 
\begin{subfigure}{1\textwidth}
  \centering
  \includegraphics[width=1\linewidth]{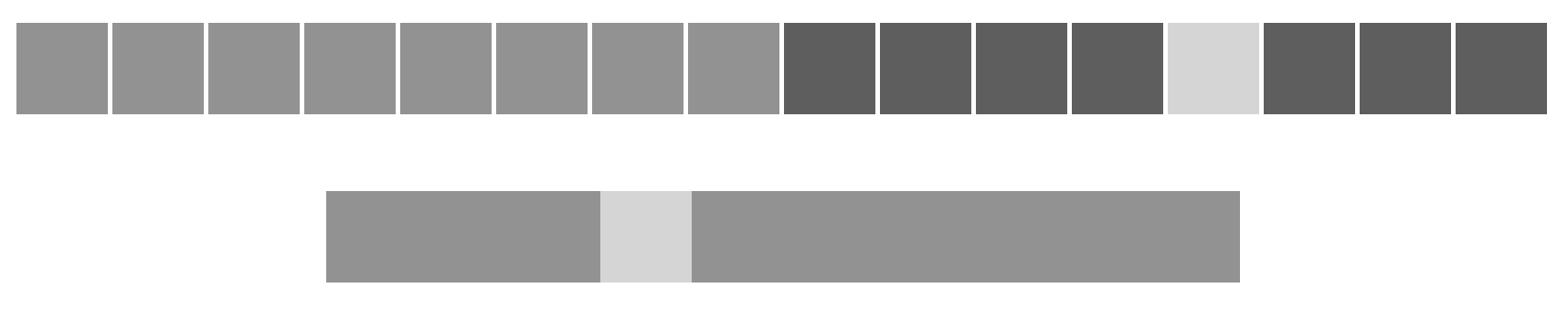}
  \caption{}
  \label{fig:colour_constancy_ex2}
\end{subfigure}
\caption{Colour constancy law. (\subref{fig:colour_constancy_ex1}) Two rows of squares are shown where in the top row two groups of squares are perceived as being grouped by their colour constancy. In the bottom row a more perplexing example is shown where a rectangular block is perceived instead of a number of adjacent squares which actually compose the row.  (\subref{fig:colour_constancy_ex2}) The same rows of squares but now with a differently shaded square placed. We immediately perceive the anomaly in the top row with respect to the darker squares, and also the anomalous square in the bottom row with respect to the homogenous connected region.
}
\label{fig:colour_constancy_ex}
\end{figure}

The colour constancy gestalt law states that if in connected regions luminance or colour does not vary strongly then the regions become unified to be perceived as a whole object with no parts inside, and in a more relaxed version two sets grouped by their colour constancy are perceived as two wholes.  In the top row of Figure \ref{fig:colour_constancy_ex}(\subref{fig:colour_constancy_ex1}) two sets of squares are naturally grouped by \emph{colour constancy} with the shade of grey---one lighter than the other. A more perplexing example is the bottom set of squares that are perceived as a single rectangular block object, as opposed to many squares in a row. In Figure \ref{fig:colour_constancy_ex}(\subref{fig:colour_constancy_ex2}) a different coloured square is introduced into both rows of the squares that are perceived immediately as anomalies with respect to the grouping.

The \emph{vicinity} or \emph{proximity} law applies in Figure \ref{fig:vicinity_ex}(\subref{fig:vicinity_ex1}) to group the black circles into two clusters that are perceived as two objects. Here, the proximity of circles within clusters is small enough with respect to the distance between the clusters to form the two groupings. Figure \ref{fig:vicinity_ex}(\subref{fig:vicinity_ex2}) adds an additional observation (black circle) that is relatively distant from both groups, and thus is perceived as an anomaly with respect to both groups. Representation by proximity in spatial distance is often used by anomaly detection algorithms.

\begin{figure}[!tbp]
\begin{subfigure}[b]{0.4 \textwidth}
  \includegraphics[width=\textwidth]{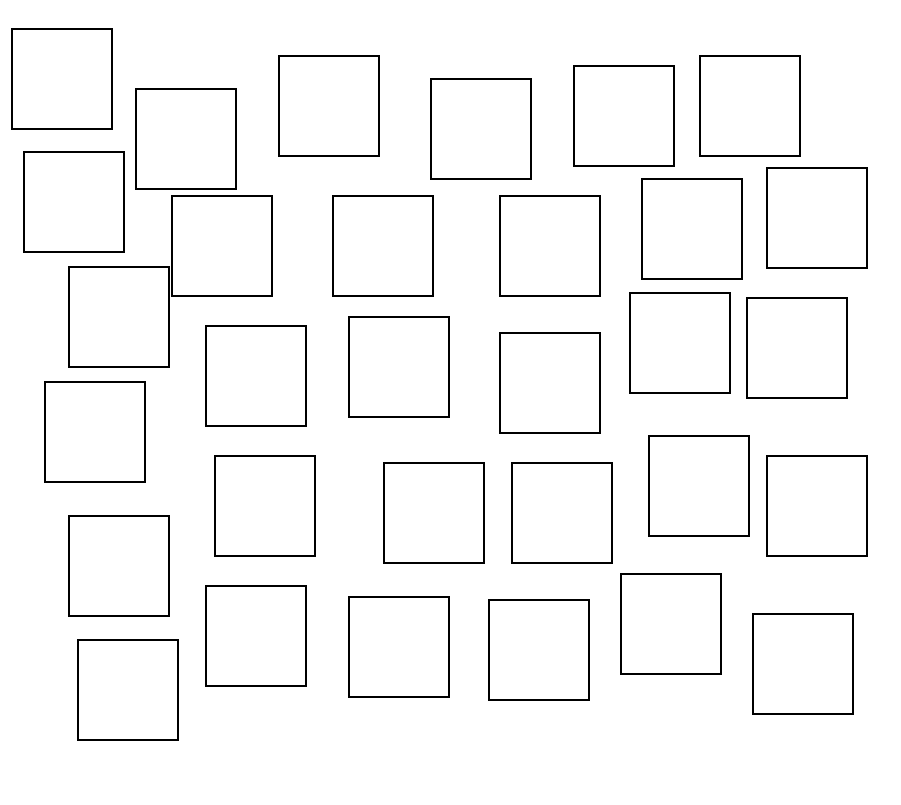}
  \caption{}
  \label{fig:similarity_ex1}
\end{subfigure}
\hfill
\begin{subfigure}[b]{0.4 \textwidth}
  \includegraphics[width=\textwidth]{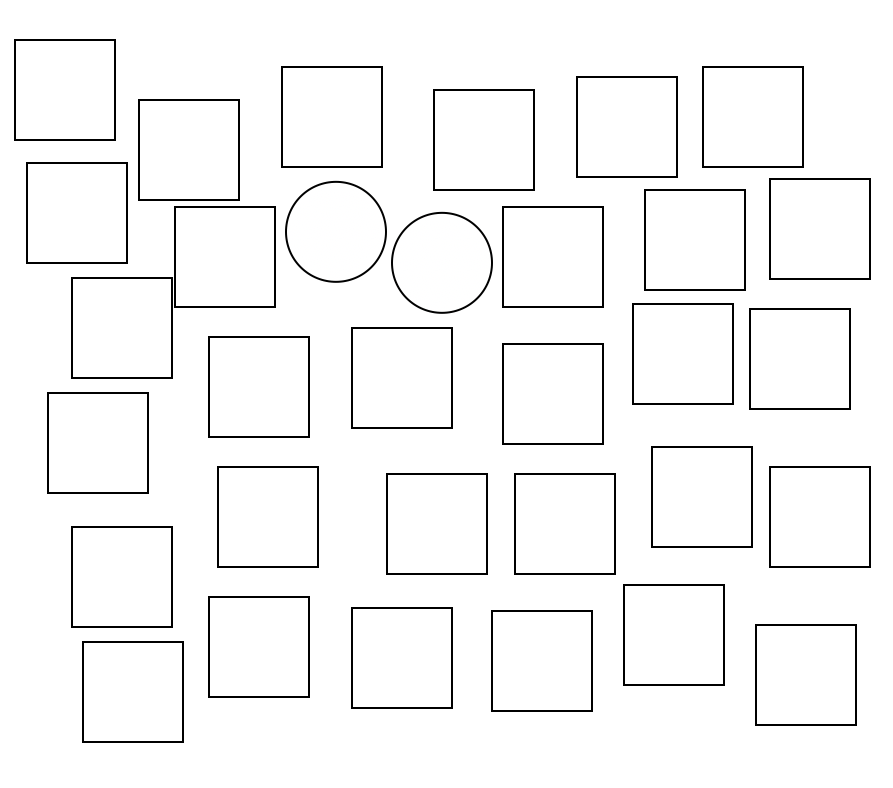}
  \caption{}
  \label{fig:similarity_ex2}
\end{subfigure}
\caption{Similarity law. (\subref{fig:similarity_ex1}) An assembly of squares is shown that are perceived as one group.  (\subref{fig:similarity_ex2}) The same arrangement but with one square in the centre replaced by two circles. These are perceived as anomalies with respect to grouping by shape, but not by colour, size or vicinity. 
}
\label{fig:similarity_ex}
\end{figure}

The \emph{similarity law} is shown in action in Figure \ref{fig:similarity_ex}(\subref{fig:similarity_ex1}) where the arrangement of squares are perceived as a group due to the fact that they are all the same with respect to measures such as shape, size, colour and proximity. Two anomalies can be detected in Figure  \ref{fig:similarity_ex}(\subref{fig:similarity_ex2}) where two circles are placed with replacement of a square towards the centre of the image. These are seen as anomalies with respect to the rest of the objects grouped by their shape, but not by colour, size or vicinity. 

Figure \ref{fig:good_continuity_ex}(\subref{fig:good_continuity_ex1}) shows an example of observations that could be grouped according to several gestalt laws---namely colour constancy, similarity in shape, local distance (vicinity) or good continuation of the pattern to perceive all the objects as belonging to one group. In Figure \ref{fig:good_continuity_ex}(\subref{fig:good_continuity_ex2}) an additional observation is added which would be considered part of the original grouping if we had considered colour constancy, similarity in shape or vicinity as the gestalt. However, when considered with respect to the good continuation grouping, which is most strongest here, it appears immediately as an anomaly. 

%
\section{Expository Examples and the Helmholtz Principle}
\label{section:motivation}
%

This section gives an exposition of examples adapted from the work by \citet*{Morel07} for providing intuition and formalism in developing the perception anomaly detection algorithm. The first is the classical \emph{number of shared birthdays in a class} problem. This example illustrates the difficulties of calculating probabilities of some simple events and why it is preferred to use expectations of the occurrence of the event to provide estimates of the probabilities. Not only will expectations be much more intuitive, easier to calculate and work with, but they also can be applied universally with some adjustment. The second example is on the perception of certain sequences in tossing a coin that shows it is not just low probability of events that are of interest, but also those that correspond to the gestalt groupings and have expectation of occurrence strictly less than $1$. Finally a general explicit statement of the Helmholtz principle and formula is given to define $\varepsilon$-meaningful events. 

\subsection{The Birthday Problem}
\label{section:birthday}
In the classical shared birthdays in a class problem we ask the following: \emph{in a class of say $n=30$ students, is it surprising to have two students share the same birthday? If not, then what about if three students share the same birthday?} 

Let us assume that their birthdays are independent and equally likely to fall on any of the $365$ days of the year. We call for $1\leq n \leq 30$, $C_n$ the number of $n$-tuples of students in the class having the same birthday. For example, $C_2$ counts the \emph{number of times} pairs of students have the same birthday: if students 1, 2 and 3 share the same birthday then we exhaustively count $(1,2), (1,3)$ and $(2,3)$, and in the case of $C_3$ we would only count how many triplets share the same birthday: (1,2,3). We also define $\mathbb{P}_n$ the probability that there is at least one $n$-tuple with the same birthday i.e. the probability of the event $C_n \geq 1$, or in other words that there is at least one group of $n$ students having the same birthday. Thus $\mathbb{P}_2$ is the probability that one or more pairs of students have the same birthday. The primary interest is in the evaluation of $\mathbb{P}_n$ and of the expectation $\mathbb{E}(C_n)$ as good indicators for the exceptionality of the event.

Given our class size of 30, and assuming independence of birthdays for all students, the probability that no two students share the same birthday is $1\times(364/365)\times(363/365)\times \dots \times(336/365)=(365\times364\times \dots \times336)/(365^{30}) \approx 0.294$. Therefore the probability that there are at least two students that share the same birthday is $\mathbb{P}_2 \approx 0.706$. In the case of $\mathbb{P}_2$ we have a straightforward calculation, however trying to approximate $\mathbb{P}_3 \approx 0.0285$ or $\mathbb{P}_4 \approx 5.4 \times 10^{-4}$ becomes complicated and for $\mathbb{P}_n$ rather counterintuitive (see \citealt[chap.~2]{Morel07} for details). However, we can approach these probabilities of the unlikeliness of shared birthdays by computing expectations instead, and by the Markov inequality, $\mathbb{P}(C_n \geq a) \leq \frac{\mathbb{E}(C_n)}{a}$ for $a>0$, we know that expectations give hints on probabilities. 

Tackling the problem through expectations, we can ask \emph{what is the expected number of pairs of students that share the same birthday in our class of 30?} Let us enumerate the students from $i=1$ to $30$ and call $E_{ij}$ the event that \emph{students $i$ and $j$ have the same birthday}. We also call $\chi_{ij} = \mathbbm{1}_{E_{ij}}$ the indicator function which is equal to $1$ if students $i$ and $j$ share the same birthday, and $0$ otherwise. We thus have $\mathbb{E}(\chi_{ij}) = \mathbb{E}(\mathbbm{1}_{E_{ij}}) = 1 \times \mathbb{P}(E_{ij}) + 0 \times \mathbb{P}(\widetilde E_{ij}) =  \mathbb{P}(E_{ij}) = 1/365$.  The expectation of the number of pairs of students having the same birthday is thus,
\begin{equation*}
\mathbb{E}(C_2) = \mathbb{E} \left( \sum_{1 \leq i \leq j \leq 30} \chi_{ij} \right) = \sum_{1 \leq i \leq j \leq 30} \mathbb{E}({\chi_{ij}})  = \frac{30 \times 29}{2} \frac{1}{365} \approx 1.192
\end{equation*}
where we have used $\sum_{1 \leq i \leq j \leq 30} = {30 \choose 2} = \frac{30!}{(30-2)!2!}$ (the number of unique pairs in a set of size $m$, subject to the commutative property $(i,j)= (j,i)$). The general formula for $C_n$, the number of $n$-tuples, follows by analogous reasoning and is given by 
\begin{equation}
\mathbb{E}(C_n) = {30 \choose n} \frac{1}{365^{n-1}}
\end{equation}

We see here how simple and intuitive the computation of $\mathbb{E}(C_n)$ is, and using this formula we have $\mathbb{E}(C_3) \approx 0.03047$ and $\mathbb{E}(C_4) \approx 5.6 \times 10^{-4}$. By the Markov inequality we only know that $\mathbb{P}_n \leq \mathbb{E}(C_n)$, however, the situation would be quite different if $\mathbb{E}(C_n)$ were small. In that case an estimate on $\mathbb{E}(C_n)$ will give us a \emph{cheap} estimate on $\mathbb{P}_n$. This is exactly what we get for $\mathbb{E}(C_3)$ and $\mathbb{E}(C_4)$ which show that indeed we have good estimates. For $n \geq 3$, this computation gives us exactly the same information as the computation of $\mathbb{P}_n$, namely the unlikeliness of $n$-tuplets. $\mathbb{E}(C_n)$ and  $\mathbb{P}_n$ differ by a small amount and give exactly the same orders of magnitude. 
The value of $\mathbb{P}_2$ tells us that it is likely to have two students who have the same birthday. $\mathbb{P}_3$ tells us that it is rare to observe triplets of students with shared birthdays, and $\mathbb{P}_4$ tells us that quadruplets are very unlikely.

\subsection{Coin Tossing Sequences}
\label{section:coin_tossing}

Let us consider the classical coin tossing examples to study the probability of tossing long runs of heads, albeit with a biased coin where the probability of heads is $(18/37)$. We can ask \emph{what is the probability that heads appears 22 times in a row?} Which is simply $(\frac{18}{37})^{22} \approx 10^{-7}$. Calculating the probability that this happens in $n$ trials can be a bit intricate. However, we can instead directly compute the expected number of occurrences of the event as $(n-(22-1)) \times (\frac{18}{37})^{22}$. If this value is $ \geq 1$, then it is likely to happen, which yields approximately $n \geq 10^7$. Less than this and we would not expect the event to occur, and would consider it unexpected. 

This example illustrates a number of important points. Firstly, the measure of exceptionality by low probability can easily apply to many different sequences of heads and tails. All possible long sequences are equally exceptional since each and every one has very low probability of occurrence. There would therefore be no surprise in an exceptional one happening, since one of the sequences must happen. Thus the comments would be incomplete if we did not also mention about human perception obeying gestalt laws. Here, most of the observations that `stand out' and are perceived as unusual belong to a small number of possible gestalts, with all other sequences taken to be usual and not to be noticed. The preceding estimates should then take into account the number of possible gestalts, not just series of heads. The limited number $N_g$ of gestalts can be estimated to include:
\begin{itemize}
\item long enough series of heads,
\item long enough series of tails,
\item long enough series of alternate heads and tails,
\item long enough series of alternate pairs head-head-tails-tails,
\item long series of alternate triples,
\item long enough series alternating one head and two tails,
\item long enough series alternating one tail and two heads,
\end{itemize}
and perhaps we could choose a few more. Thus, if we calculate the event for \emph{any of these gestalts is observed}, we simply multiply the former expected number of occurrences by $N_g$ and the conclusions remain valid.

A second important point is that modelling the problem with appropriate random variables and expectations enables us to have a natural threshold parameter value of $1$. Here, if the expectation of the event occurring is $\geq 1$, then its occurrence is not surprising, since it is expected to occur at least once. Conversely, if the expectation is $<1$, then the event is deemed unexpected and significant as it is not expected to occur even once. This removes the problem of choosing probability thresholds and is crucial in deriving a parameter-free algorithm that first separates between unexpected and expected events, and then the actual expected value scores can be used to rank the events---the smaller the value, the more unexpected is the event.

A third important point is that it is not enough to just say that a specific a priori grouping is perceived if it has low probability to occur. We also need to complete the analysis by considering all of the events of the same kind that could occur. If there are many of them, it is likely that one of them will occur and hence the event occurring would not be significant. In the earlier \emph{birthday problem} example we saw that in a class of $30$ students, it would be rare and unexpected for three students to share the same birthday. However, if we increase the class size then the expectation of triplets occurring increases, and at a class size of $100$ students this would not be an unexpected event. Similarly, in computer vision tasks the observation of some geometric event such as \emph{these two lines are parallel} can occur with low probability, however, we need to also consider all possible groups of parallel lines in the considered image. If the image contains many lines, then it is likely that a pair of them will be quite parallel, and their detection would not be unexpected or significant. Only if the expected number of such pairs, given the total number of lines, is $<1$, can one decide that the observed parallelism makes sense. Returning to the coin tossing example, we have calculated the order of the number of trials needed to be carried out to make the expectation of the event of $22$ consecutive heads likely. Thus the exceptionality of this event occurring is based on not just low probability, but also the total number of trials $n$ that are observed. 

\subsection{The Helmholtz Principle and $\varepsilon$-meaningful Events}
\label{section:helmholtz}

The Helmholtz principle is for understanding human perception and can be stated in the following generic way given by \citet{Morel07}: assume there are atomic objects $O_1, O_2, ..., O_N$ and that $k$ of them, $O_1,..., O_k$, have a common feature (same colour, same orientation, position, same birthday, etc.) that has been independently, randomly and uniformly distributed over all $N$ objects. We can then ask if this common feature is happening just by chance or if it is significant enough to group $O_1, ..., O_k$? When we observe a random realisation of this uniform process we want to know if the grouping is likely; if not, this proves \emph{a contrario} that a grouping process---a gestalt---is at play. This principle, together with the earlier expository examples and discussions, leads us to the following definition.

\begin{definition}[$\varepsilon$-meaningful event]
We say that an event is $\varepsilon$-meaningful if the expectation of the number of occurrences of this event is less than $\varepsilon$ under the \emph{a contrario} random assumption. When $\varepsilon \leq 1$, we simply say that the event is meaningful.
\end{definition}
The above definition is very generic and must be accompanied by suitable perceptually relevant events and adequate \emph{a-contrario} assumptions that numerical qualities of objects are independently and uniformly distributed. If the Helmholtz principle is true, we perceive events if and only if they are meaningful in the sense of the preceding definition. Now, given that the probability of an object $O_i$ to have a quality is equal to $p$ (in the birthday example we had $p=\frac{1}{365}$) under the independence assumption, the probability that \emph{at least $k$} objects of the observed $N$ have this quality is given by
\begin{equation*}
B(N,k,p) = \sum_{i=k}^{N} {N \choose i}p^{i}(1-p)^{N-i}
\end{equation*}
i.e. the tail of the binomial distribution. To get an upper bound for the expectation of the number of events happening by pure chance, one simply multiplies the above probability by the number of tests performed. This number of tests $N_{conf}$ corresponds to the number of different possible configurations one could have for the searched gestalt. Thus, a considered event will be defined as $\varepsilon$-meaningful  if,
\begin{equation*}
N_{conf} \times B(N,k,p) < \varepsilon
\end{equation*}
and the smaller this is, the more meaningful the event is. 

%
\section{Unsupervised Parameter-free Anomaly Detection}
\label{section:algorithm_description}
%

This section begins with a description of a cyber security problem in identifying suspicious levels of authentication failures. The problem initiated the development of the perception algorithm which is later described using the mathematical treatment of gestalt groupings and the Helmholtz principle. This gives an elegant solution that is unsupervised, efficient and parameter-free. Thus we do not require labelled data, nor do we leave to the users the difficult task of choosing critical algorithm parameters or their tuning (which can potentially bias results). In the latter parts of this section a generalisation of the algorithm with some practical adjustments is provided to detect anomalies in univariate data. This is followed by an extension for multivariate data.

\subsection{Authentication Failures in Cyber Security Systems}

\begin{figure}
\centering
  \includegraphics[width=1\linewidth]{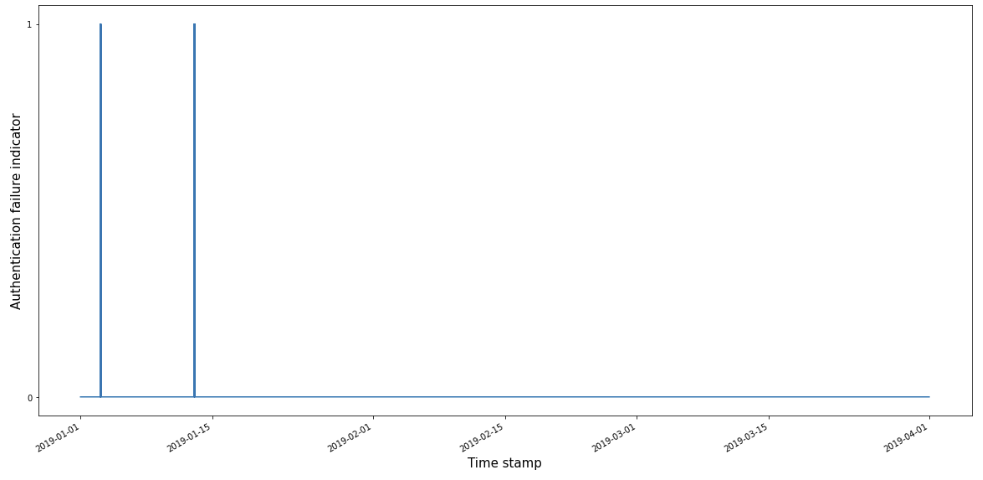}
            \\[10pt]
  \caption{A binary $\{0,1\}$ valued line graph is shown of the authentication failures recorded in event logs of a security monitoring system over time. Largely no authentication failures occur, however, two noticeable bursts are observed over a few hours indicating suspicious behaviour.}
  \label{fig:authentication_failures_over_time}
\end{figure}

The following problem from the cyber security domain inspired the perception anomaly detection algorithm: I was given streaming event logs from various components of a security monitoring system. One of the features that can be obtained from the logs is the occurrence of authentication failures which, in normal circumstances should not occur, although a relatively few in number could occur unsuspiciously within a given window of events (corresponding to cases such as incorrect passwords input by a user). The data can be represented as a binary stream, with $0$ corresponding to no occurrence of an authentication failure, and $1$ that a failure has occurred. Such streams will be called \emph{indicating data} or simply \emph{indicators} because they inform whether or not a feature of prior interest has occurred. A sample of the data was collected over a three month period of time and contains on the order of half a million events. Figure \ref{fig:authentication_failures_over_time} graphs the data over time. Except for two relatively short periods of time where a burst of authentication failures are logged, all other events do not record any authentication failures. The objective was to detect such occurrences which could be attributed to potential system intrusions---such as brute force attacks or denial of service---or a malfunctioning system by finding and reporting anomalous events that correspond to suspicious levels of indicating data. It was also required to rank the suspicious events in order of priority for further investigation by human analysts; such a ranking being secondary to uncovering the anomalous events.

With the events ordered by time, aggregate counts of indicators using adjacent fixed size windows can be used as a simple method to generate numerical values to be checked for anomalies. It is initially assumed that the set of training data can be processed in batch. Any new indicating data that arrives can be similarly processed with adjacent (or sliding) windows of the fixed size. For each window we thus have a count of how many authentication failures that have occurred and the values can be thought of as a stream of numbers. For simplicity I assume that the chosen integer window length $L$ divides into the number of events in the training data set, $N$, without remainder to give the number of windows $W$. For example, given a stream of indicating data, $D=[0,0,0,0, 1,1,1,1, 1,1,0,1, 0,0,0,0, 0,0,0,0]$, of length $N=|D|=20$, using $W=5$ adjacent windows of length $L=4$ we obtain the window values $V=[0,4,3,0,0]$. 

In the actual cyber-security problem the number of windows $W$ and hence the list $V$ is very big, with almost all values being $0$ and corresponding to no authentication failures in a given window, and the rest take values up to a maximum of $L$. A person viewing these values with no background knowledge would, according to our a-contrario model of assuming that the numbers are distributed independently and uniformly at random within the range $[0,L]$, naturally consider the large number of $0$'s as a meaningful group (according to, say, similarity or vicinity), since according to the Helmholtz principle, we perceive what is unlikely to have happened by chance. 

However, for our purposes we are interested in detecting the anomalies with respect to this grouping, and which correspond to the relatively few integers that are much larger in value. Indeed, a person would also likely notice the relatively few non-zero values as being anomalous with respect to the number of zero values. How large an integer must be to be considered an anomaly is an important question. One can imagine the analogy of pulling out a ball from amongst a group of identical  balls, where at some point the ball pulled out is perceived as an outlier when compared to the group. Similarly, in order to decide which integers in the list $V$ are anomalies the method of detecting meaningful events using Helmholtz principle is used, but requires a complementary and perhaps opposite measure to the grouping by 0's: Consider summing the total number of indicators across the training data set to find $S$ (for our example data $[0,4,3,0,0]$ this would give $S=7$), we can then ask: \emph{given the total sum of integers $S$, windows $W$ each of length $L$, and $1\leq n \leq L \leq S$, is the occurrence of $n$ authentication failures in a given window unexpected or not?}  

\begin{figure}
\centering
  \includegraphics[width=1\linewidth]{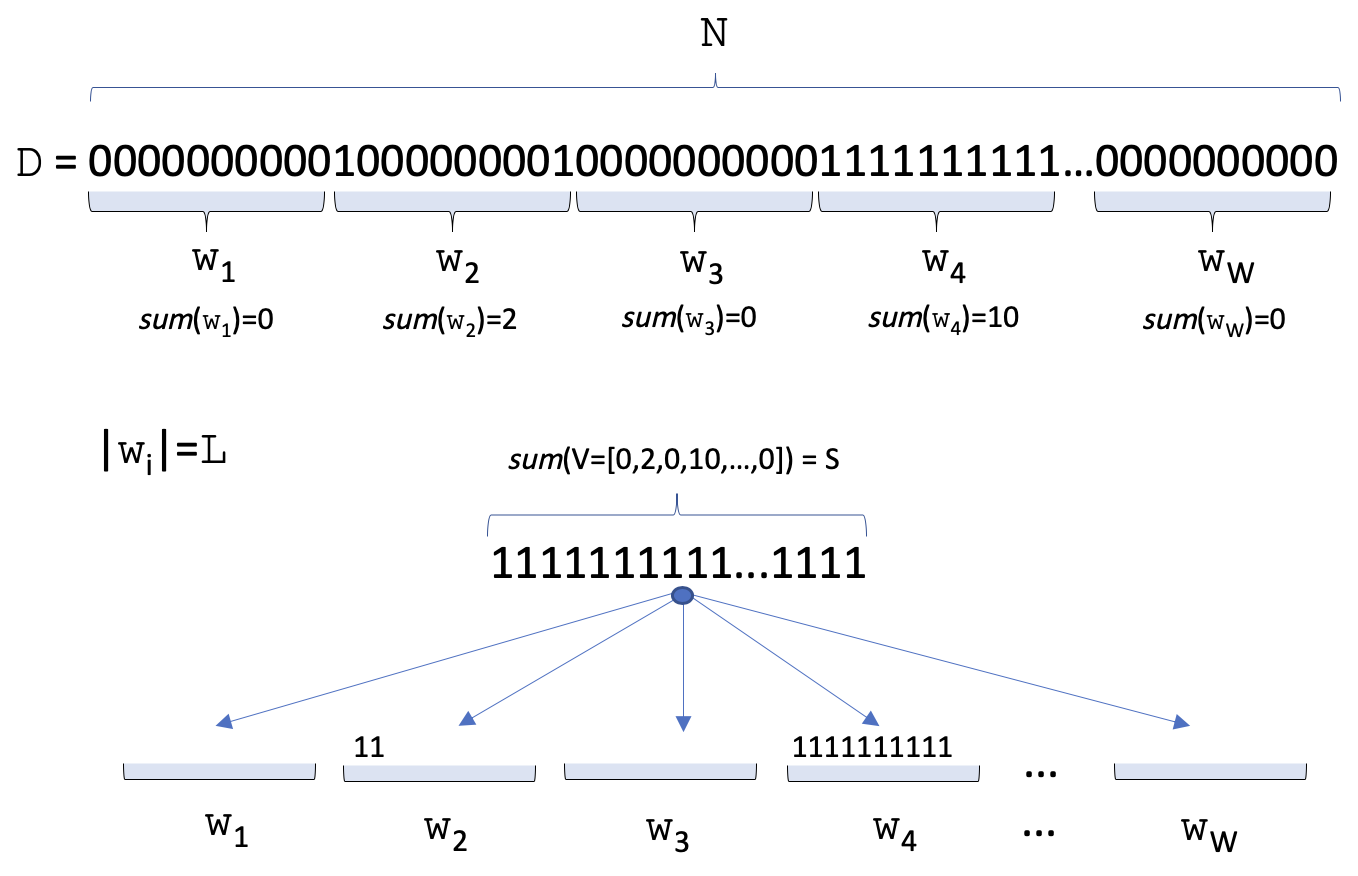}
  \\[10pt]
  \caption{A stream of indicators $D$ being represented by windows of length $L$ that count how many nonzero indicators occur in the given window. The a-contrario assumption is that the total sum of indicators is uniformly randomly and independently distributed over the $W$ windows. Thus, given the realisation of the indicator stream, we can ask if it is unexpected to have $n$ nonzero indicators in a given window $w_i$ under the a-contrario assumption.}
  \label{fig:distributing_indicators}
\end{figure}

To answer this question the analysis of the birthday problem in section \ref{section:birthday} is generalised and Figure \ref{fig:distributing_indicators}  is provided to aid the explanation. Given some indicating training data stream $D$, of length $N$, let $W$, enumerated as $w_1, w_2, \dots w_W$, be the number of windows where each window is of equal length $L$, and assume that $N$ is wholly divisible by $L$. Each $w_i$ views a set of $L$ consecutive indicators from $D$, and we can sum all the indicators in each window to give $V=[v_1, v_2, \dots v_W]=[sum(w_1), sum(w_2), \dots sum(w_W)]$. Now taking the sum of \emph{all} the indicators $S=sum(V)$, the a-contrario assumption and model is that all the nonzero indicators ($1's$ of total sum $S$) are uniformly and independently randomly distributed across all the windows $w_i$. The gestalt quality of interest is the unexpected number of nonzero indicators that appear in a window under the a-contrario assumption. 

As we saw for the birthday problem in section \ref{section:birthday}, we can use $\mathbb{E}(C_n)$ for measuring the unexpectedness of the occurrence of an $n$-tuple of the nonzero indicators, $1's$ from $S$, distributed over the windows $w_i$, as a measure of how anomalous the number of elements in a particular window are. (The analogy with the birthday problem is that $S$ can be thought of as the total number of students in a class, with each integer $1$ from $S$ representing a student, and $W$ the number of possible birthdays, with the a-contrario assumption that the students are equally likely to have a birthday on any day $w_i$. Thus in a class of $S$ students we would ask \emph{is it unexpected to have an $n$-tuple of students share the same birthday $w_i$?}) We also saw from section \ref{section:coin_tossing} that we are interested in events where the expected value is $<1$, which acts as a natural threshold for whether we expect the event to occur given the data parameters. Thus, the general formula for measuring the unexpectedness of each $n$-tuple and flagging an anomaly in $V$ is

 \begin{equation}
 \label{equation:anomaly_detector}
\mathbb{E}(C_n) = {S \choose n} \frac{1}{W^{n-1}} <1
\end{equation}
Note the simplicity of the formula, as expected to be given by a good principle. This calculation gives an immediate binary output on the training data to class each window (value from V) as anomalous or benign---hence all the events within the window as well. There are no parameters left to the user to specify. The use of expectations handles the probability thresholding problem to immediately arrive at the anomalous events, which can subsequently be ranked with smaller expectations corresponding to more meaningful events. Under this construction the previously meaningful group of \emph{$0$} values has become the expected norm, and the outlying nonzero values have become the unexpected and meaningful. 

Adjacent windows that exactly fit into the length of the indicating data stream have been used to simply the exposition. In practice this has also shown to work well with zero padding on the edge window case when fitting the model to the training data. However, when testing the training data for anomalies, it is preferred to use a sliding window in the detection step as it enables testing all the windows of length $L$ of the data stream. Thus, for each sliding window the number of nonzero indicators, $n$, is counted and checked to see if $\mathbb{E}(C_n)<1$. The test on new incoming data is also a simple matter of summing the number of indicators in the new window and testing its expectation accordingly. Alternatively sliding windows can be used as each indicator in the steam arrives. The extension to incorporate a new data window into the training data is also fast and simple; increment $W$ by one and add the sum of the new indicators to $S$ and then re-compute all required $\mathbb{E}(C_n)$. 

\subsection{The Perception Algorithm}
\label{subsection:Perception_algorithm}

The detection of anomalous counts of authentication failures in windows of streaming data can be generalised to give the parameter-free perception algorithm. Given a set of nonnegative integer numbers, consider each number having been generated by counting the number of nonzero indicators in adjacent windows of the same length in a data stream. Using the running example set of numbers $V=[0,4,3,0,0]$, this could be thought of as being generated by the stream $[0,0,0,0, 1,1,1,1, 1,1,0,1, 0,0,0,0, 0,0,0,0]$. The total number of items in $V$ gives us the number of windows $W=5$ and the sum of the numbers gives $S=7$. Under the a-contrario assumption of the independent and uniformly randomly distributed nonzero indicators, the unexpectedness of each integer $n$ in $V$ can be calculated by plugging the parameters $S, W$ and $n$, obtained directly form the data itself, into Equation \eqref{equation:anomaly_detector}. Note that the length $L$ of each window is not required as we only need to assume that nonzero indicators occur with equal probability in any given window. 

The computation of Equation \eqref{equation:anomaly_detector} requires care in practice due to the large values of the parameters $S, W$ and $n$ in computing the binomial coefficient. Furthermore, taking logs of both sides of the inequality transforms it to a unified format for giving asymptotic estimates of the general formula:

 \begin{equation}
 \label{equation:anomaly_detector_log}
-\frac{1}{S} \left( log{S\choose n}  - (n-1) log(W) \right) > 0
\end{equation}
Under this transformation the anomalous events are now identified as those whose calculation outputs are greater than $0$; the greater the value the more anomalous or unexpected is the event. The computational difficulty here lies in computing $log {S \choose n}$ due to computing overflow issues and is expanded out to

 \begin{equation*}
log(S!) - log((S-n)!) - log(n!)
\end{equation*}
where each of these log factorials is estimated efficiently using Stirling's approximation

 \begin{equation*}
log(m!) \approx mlog(m) - m + (0.5 \times log(m)) + (0.5 \times log(2\pi))
\end{equation*}

The input data requires preprocessing to include handling both positive and negative numbers to a specified number of significant digits. This is done by first rounding all the data values to the required accuracy and multiplying them by $10$ to the exponent of the largest number of decimal places. This transforms all the values into integers. For example, given an array of numbers $Q=[-0.1, -1.46, 1.2, 1.35, 2.678, 2.10293, 10]$ and accuracy of $2$ decimal places, this would convert to $[-10, -146, 120, 135, 268, 210, 1000]$. The default accuracy of $4$ is empirically sufficient unless the data has unusually small values or differences, in which case some inspection is required to choose the appropriate number of significant digits. The next transformation required is to take the magnitude of the distance of the data to the median for the subsequent gestalt grouping by proximity. The median is computed to the nearest integer and found in the current example to be $135$. The median is used because it is a robust estimator (less affected by outliers) and the assumption that human judgment uses it as a point of natural centrality. This yields the transformed values $[125, 281,  15,   0, 133,  75, 865]$, which are used to compute the parameters of the fitted model by taking the sum and length of the array: $S=1494$ and  $W=7$. Note the simplicity of the model fit, and the obvious speed at which this can be done. In the testing phase each value of the test array is similarly transformed (the median of the training phase is used), and the unexpectedness calculated using Equation \eqref{equation:anomaly_detector_log} where any computed score $>0$ is considered to be an anomaly. Once again the test phase is also extremely fast to compute for all values. In current example the value $10$ in $Q$ is calculated to be an anomaly. The full perception training and testing algorithms for the detection of global anomalies in univariate numerical data is detailed by Algorithms (\ref{algorithm:PerceptionTrain}) and (\ref{algorithm:PerceptionTest})---albeit generalised for multivariate data.

\subsection{Generalisation to Multivariate Data}
\label{subsection:multivariate_data_algorithm}

The perception algorithm can be generalised for application to multivariate data where there is more than one feature of interest. It is assumed that there is \emph{only one} normal grouping of the data that largely forms a concentrated mass of points and that anomalies are global in nature---lying on the fringes of the data or further out still. While this may appear overly restrictive, many anomaly detection problems are naturally of this type. 

Intuitively, anomalies are considered to be those points in multidimensional space that lie an unusually large distance from the central grouping---represented best by the median---compared to the rest of the data points. Thus each feature column is first transformed to have the same standard deviation so that each is on a similar scale and no one feature dominates distance calculations. This is an important preprocessing step where the selection of the standardising method can have a significant impact on the results. Then the Euclidean distance of every data point to the multidimensional median is calculated. This yields a one-dimensional array which is transformed to integers as before in section \ref{subsection:Perception_algorithm} (rounded and multiplied by $10$ to the exponent of the largest number of decimal places, and the distance to its median derived), and the sum of their values, $S$, and length of the array, $W$, computed. The unexpectedness score of each value of the array (which correspond to example rows of the input data) is then calculated to report anomalies wherever the score is $>0$ using Equation \eqref{equation:anomaly_detector_log}.  

This multivariate generalisation of the algorithm collapses to the univariate version when the data is one-dimensional. Henceforth the perception algorithm will refer to this multivariate version of the algorithm. The training and test phases for a data set are given by Algorithms (\ref{algorithm:PerceptionTrain}) and (\ref{algorithm:PerceptionTest}). 

\begin{algorithm}
\SetAlgoLined
\caption{PerceptionTrain}\label{algorithm:PerceptionTrain}
\DontPrintSemicolon
  \KwInput{	  	
  	$X$  - input $l \times m$ dimensional array of numbers \newline
  	$acc$ - decimal place accuracy of input data (default 4) \newline
  	$metric$ - distance metric to be used (default Euclidean) \newline
	$\mu$ - mean of each feature of X  \newline
	$\sigma$ - standard deviation of feature of X
}
\BlankLine
  \KwOutput{  
  	$S$ - sum of indicators \newline
    	$W$ - number of windows \newline
    	$multiDimMed$ - median values of each feature of X \newline
    	$med$ - median of one-dimensional data 
}
  
\BlankLine
\If{$dimension(X) > 1$} {
	$X_s, \mu, \sigma \leftarrow standardise(X)$\; 
	$multiDimMed \leftarrow median(X_s, axis=column)$ \; 
	$X_d \leftarrow distance(X_s, multiDimMed, metric)$\;        	
	$X \gets X_d$
}
\BlankLine
$X_g \leftarrow roundAndMakeInteger(X, acc)$	\;		
$med \leftarrow round(median(X_g))$\;
$X_f \leftarrow |X_g - med|$\;
$S \leftarrow sum(X_f)$\;					
 $W \leftarrow length(X_f)$\;
 
\end{algorithm}

\begin{algorithm}
\SetAlgoLined
\caption{PerceptionTest}
\label{algorithm:PerceptionTest}
\DontPrintSemicolon

  \KwInput{	
  	$X$ - input $l \times m$ dimensional array of numbers to test \newline
  	$S, W, multiDimMed, med, acc, metric, \mu, \sigma$ - from training stage
}
\BlankLine
  \KwOutput{  
  	$Z$ - array of prediction scores for each example row in $X$ \newline
    	$Y’$ - binary array of predictions for $X$ (1: anomaly, 0: normal) 
}
\BlankLine

\If {$dimension(X) > 1$} {
	$X_s \leftarrow standardise_{\mu, \sigma}(X)$\;  
        	$X_d \leftarrow distance(X_s, multiDimMed, metric)$\;        	
	$X \gets X_d$
}
\BlankLine
	$X_g \gets roundAndMakeInteger(X, acc)$\;			
	$X_f \gets |X_g - med|$
\BlankLine
	Initialise $Y'$ to zero array of length $l$\;
  	\For{$i = 0$ to $l-1$}{
		$n \gets X_f[i]$\;
		$Z[i] \leftarrow -1/S \times (log{S \choose n} - (n - 1) \times log(W))$\;	
		\If {$Z[i] > 0$}{
			$Y’[i] \leftarrow 1$ 		
		}
	}
	
\end{algorithm}

%
\section{Evaluation and Results}
\label{section:implementation}
%

This section reports on the performance of the perception algorithm on a variety of data sets and compares it against other anomaly detection algorithms. However, a thorough comparison is fraught with difficulties and avoided for the following reasons: 

\begin{enumerate}
\item \emph{Level of supervision}. Some algorithms operate in supervised mode where availability of labelled data is assumed. Others are semi-supervised where usually only normal data labels are available. The most common scenario is to have an unsupervised setting where it is assumed no data labels are available except perhaps for a few that are usually reserved for validation purposes. 
\item \emph{Parameter choice}. Most---if not all---algorithms have one or more influential parameters that are left to the user to specify and can be difficult to choose in unsupervised learning tasks. For example, the value $k$ in nearest neighbour algorithms or the percentage of anomalies assumed to be present in the data set. Thus, the `unsupervised learning algorithms' actually resort to requiring their users to supply data dependent parameters. 
\item \emph{Data distribution}. The same algorithm that works well with one assumed distribution of data may fail on another, e.g. symmetric verses skewed. 
\item \emph{Data properties}. On real-world data sets the number of feature dimensions, type of data (numerical, nominal, ordinal, etc.) or the number of observations can all affect the performance and application, with different algorithms giving different results depending on these properties.
\item \emph{Anomaly type}. The unknown anomalies may be global or local in nature. Some algorithms are better on the former than the latter, or vice-versa; others aim to handle both types.
\item \emph{Criteria for success}. The criteria for a successful method can vary between applications. For example, in one case it may be desired to have a method that performs well on both Gaussian and non-Gaussian distributed data, while in another case the speed of the algorithm may be given more relative weight than precision or recall. 
\item \emph{Availability of ground truth}. Real-world anomaly data sets with highly accurate labels are difficult to obtain and in large data sets it may be infeasible to know and label all instances of anomalies. Instead labelled data sets that are used for classification problems are usually modified. However, these data sets may not accurately reflect real-world anomaly detection problems, and hence the performance of algorithms. 
\end{enumerate}

The reasons above highlight the fact that anomaly detection in practice requires a thorough understanding of the problem space and criteria for success. Rather than using an algorithm blindly, some data exploration is advised in order to ascertain whether the problem data sets (current and future) are expected to at least approximate the assumptions of the algorithms being considered. One of the primary concerns here is that applying algorithms to different data sets can and do produce different results. Indeed, even the same algorithm applied to the same two data sets, but where one has been pre-processed differently to the other,  can result in very different results (e.g., when making anomaly detection data sets from labelled classification data). These issues make the comparison difficult, but a general picture can emerge where we observe particular algorithms performing better on average or on certain categories of data. Hence, in the present work the perception algorithm is tested on a limited number of data sets and types, and against other algorithms with their default parameters selected. The results highlight where it performs well, and where it may fail in order to gain empirical understanding of its application and the key benefits of the approach. 

\subsection{Univariate Data Results}

Univariate data is defined as having only one feature or dimension of interest. The evaluation is done against the following statistical methods: 
\begin{itemize}
\item z-score; a threshold value of $3$ standard deviations from the mean is used.
\item modified z-score; a threshold of $3.5$ mean absolute deviations (MAD) from the median is used.
\item Tukey's box-plot (IQR) with the frequently used $1.5 \times$IQR for making cutoff points.
\end{itemize}
These methods are the nearest we come to comparing parameter-free algorithms since the threshold values are so commonly cited in outlier detection literature. The methods all assume that the data is Gaussian distributed and numerical and thus it could be argued that any comparison on non-Gaussian data sets is unfair; however, good performance on both distributions is a clear advantage. Indeed, in practice they can still be used to detect anomalies even if this assumption is unsatisfied. The detection is also restricted to global anomalies in numerical data; leaving the finding of local anomalies or anomalies in nominal data to future work. 

The choice of metric to evaluate the algorithms is also an important consideration. In the case of small univariate data samples the anomalies are few and easily observed by the human eye. The expectation here is that the algorithms should find all the obvious anomalies and perhaps propose others in fringe cases. For larger univariate data sets where the selection of anomalies is more subjective or domain specific, I simply show and subjectively compare the results; this is due to being concerned with point anomaly detection and because the data is not labelled as such. The lack of an objective metric does not prevent from meaningfully demonstrating the perception algorithm. 

\subsubsection{Small Data Sets} \label{small_univariate_datasets}
The first data sets are small manually crafted ones that we can intuitively grasp to perceive the anomalies. The study on small data sets is not only a natural place to start, but is also of practical importance since although humans can readily perform anomaly detection on such data, automated methods are desired that can scale with the number of data sets and metrics. The first is a small hypothetical data set from \cite{iglewicz1993detect}:
\begin{gather*}
2.1, 2.6, 2.4, 2.5, 2.3, 2.1, 2.3, 2.6, 8.2, 8.3
\end{gather*}

\begin{figure}
\centering
  \includegraphics[width=1\linewidth]{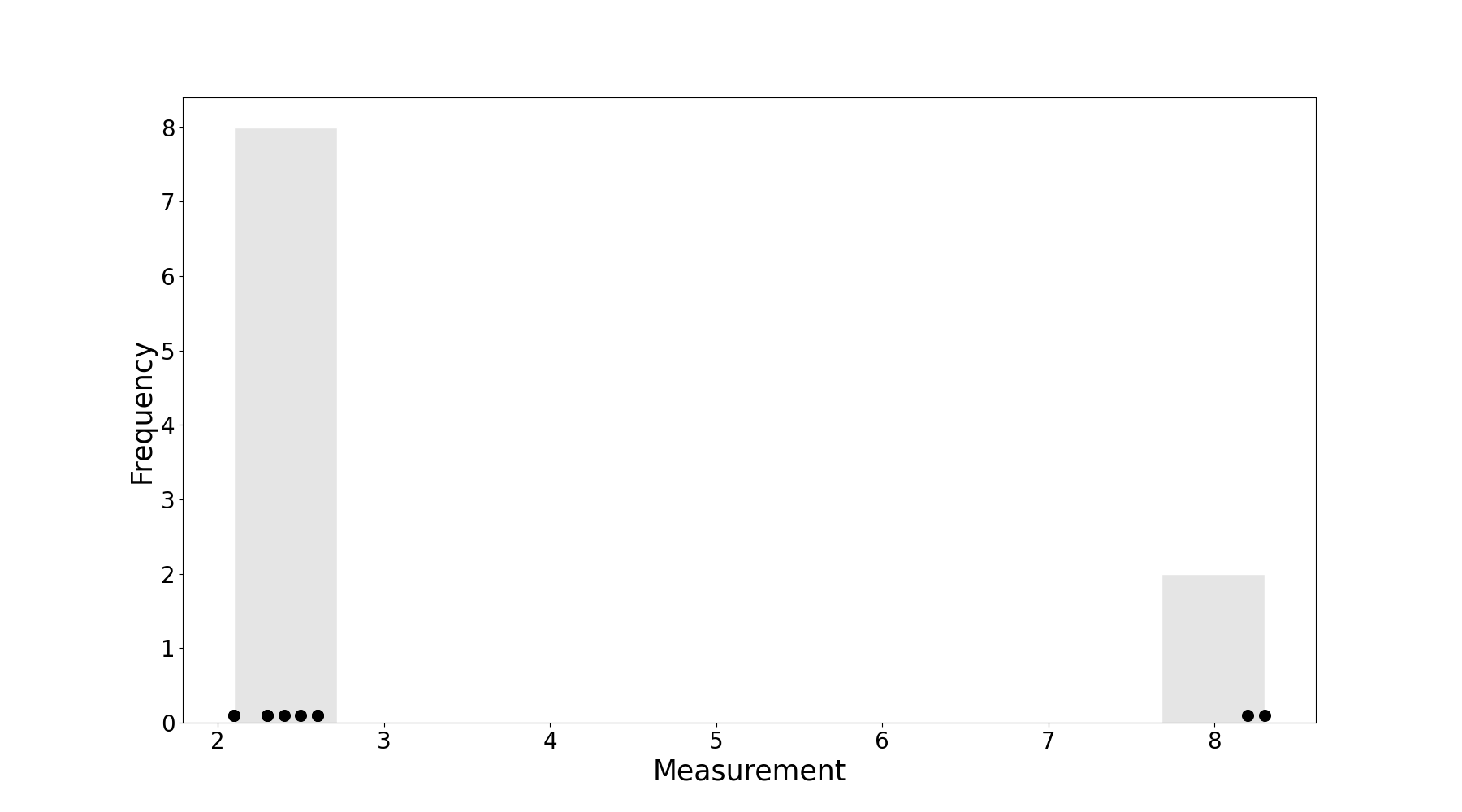}
  \caption{Histogram and scatter plot showing the hypothetical data set distribution from \cite{iglewicz1993detect}. The majority of the data form a grouping on the left and there are two outliers to the right. Anomalies detected: z-score: None; modified z-score, perception, IQR: $\{8.2, 8.3\}$.  
  }
  \label{fig:BorisHypo1}
\end{figure}

The histogram and scatter plots---two methods that can be used to visualise data sets and highlight the anomalous observations---are shown in Figure \ref{fig:BorisHypo1}. We perceive two values as anomalies due to them being relatively larger than the rest. The modified z-score, IQR and the perception algorithm detect the two anomalies $\{8.2, 8.3\}$. However, the often used z-score method fails to find any anomalies. This example is used by \citet{iglewicz1993detect} to illustrate why the commonly used z-score is not always a good outlier detector for small data sets. Furthermore, they demonstrate that because the mean and standard deviation are heavily affected by the presence of outliers, the method is highly sensitive to the presence of the actual observations it is being used to detect. 

\begin{figure}
\centering
  \includegraphics[width=1\linewidth]{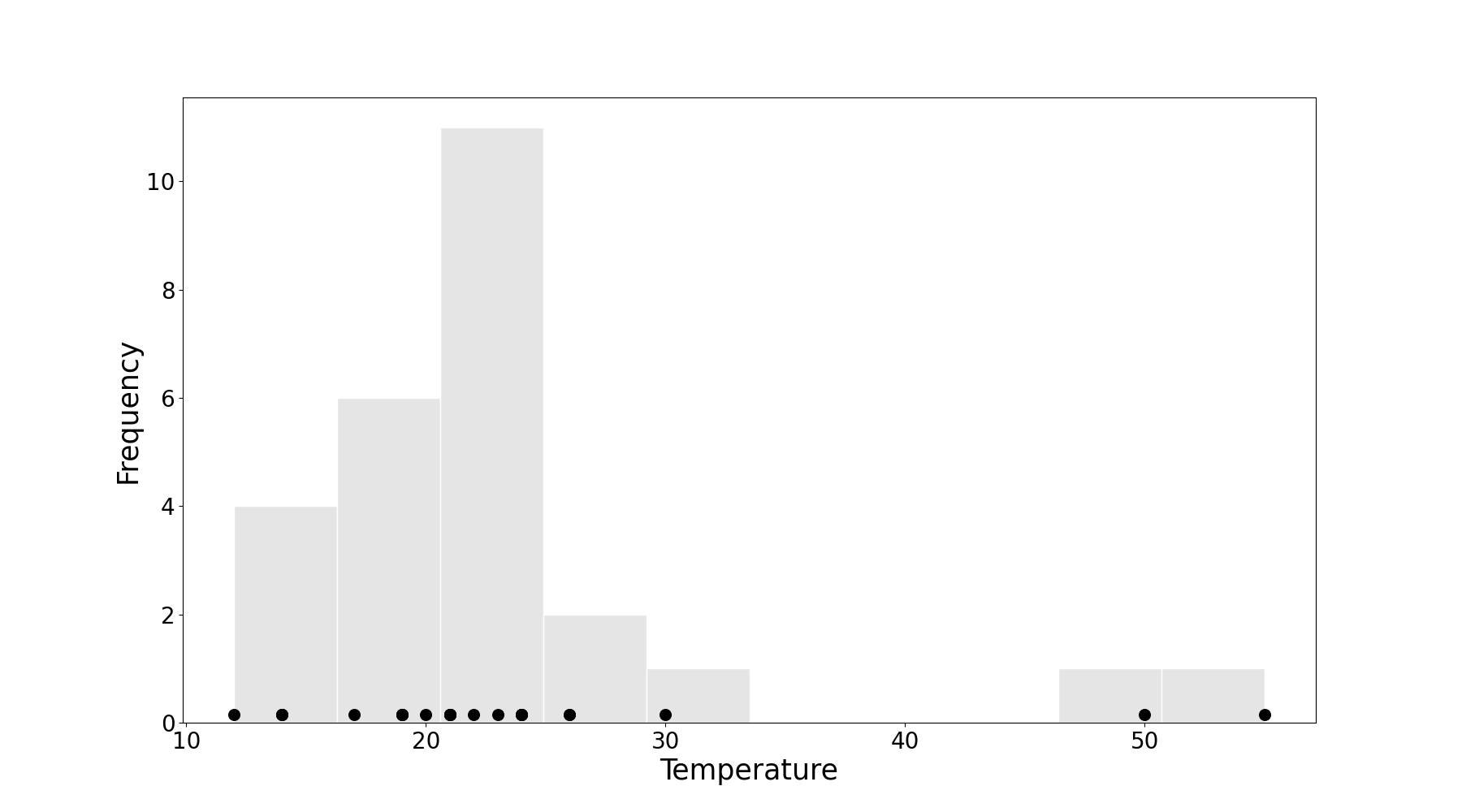}
  \caption{Distribution of temperature data with two artificially placed outliers. The majority of the data form a grouping on the left and there are two outliers to the right. Anomalies detected: z-score: $\{55\}$; modified z-score, perception, IQR: $\{50, 55\}$.  
}
  \label{fig:temperatures1}
\end{figure}

The next example is a data set of temperatures recorded in degrees celsius throughout the day with two artificially introduced outliers: 
\begin{gather*}
12, 14, 14, 14, 17, 19, 19, 19, 19, 20, 21, 21, 21, 21, \\
21, 22, 23, 24, 24, 24, 24, 26, 26, 30, 50, 55
\end{gather*}
The histogram and scatter plots are shown in Figure \ref{fig:temperatures1} including the two high valued outliers. The z-score method only finds the higher valued outlier of $\{55\}$ and fails to find the nearby value of $\{50\}$. This is a classic example of the masking phenomenon in anomaly detection. Here, the outliers affect the z-score thresholds to introduce false negatives as their existence pull the boundary away from the normal observations. The other three methods detect the two anomalies correctly. 

\begin{figure}
\centering
  \includegraphics[width=1\linewidth]{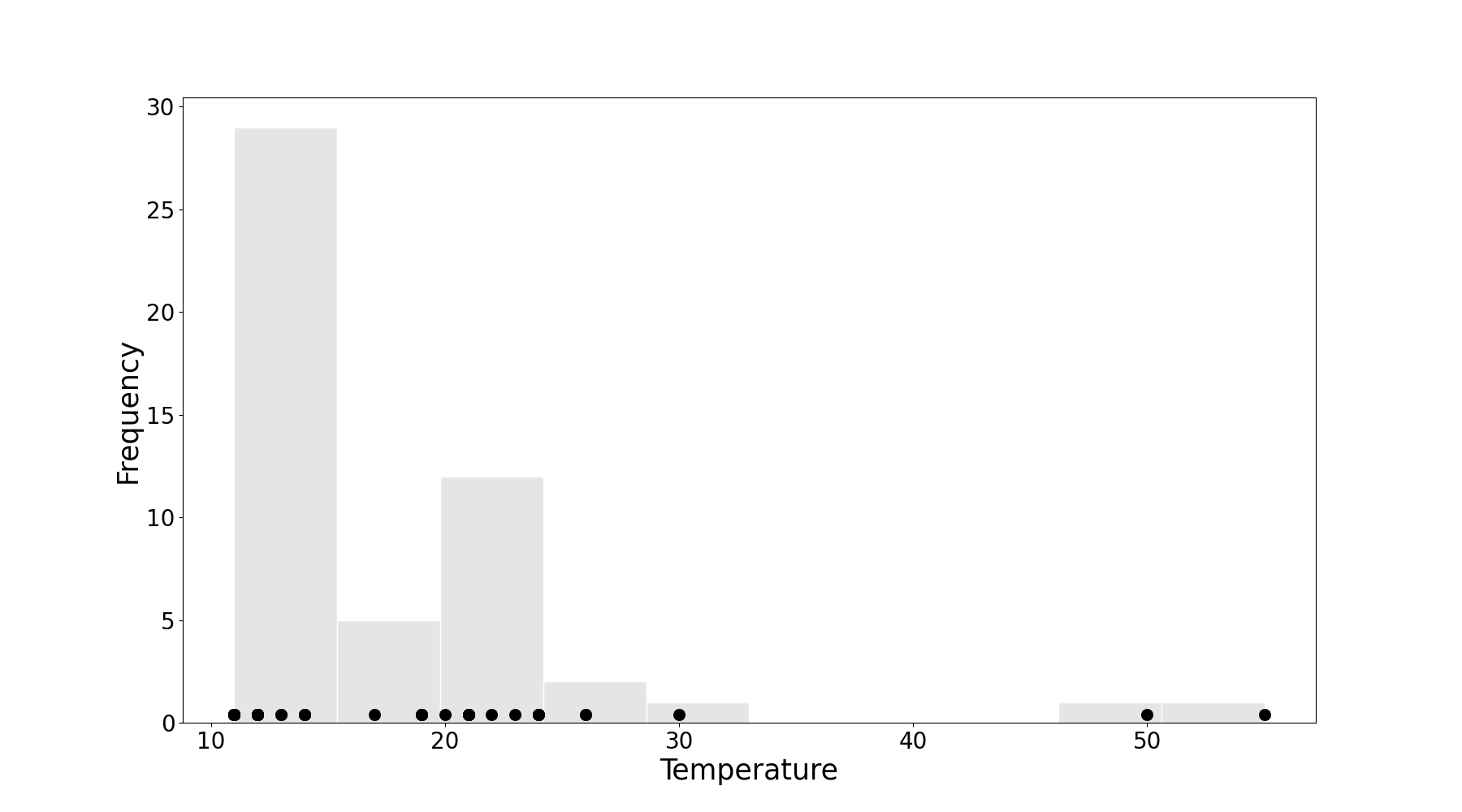}
  \caption{Distribution of temperature data with outliers and additional normal points. The majority of the data form a grouping on the left and there are two outliers to the right. Anomalies detected: z-score, perception, IQR: $\{50, 55\}$; modified z-score: $\{24, 24, 24, 24, 26, 26, 30, 50, 55\}$.  
}
  \label{fig:temperatures2}
\end{figure}

The next example is the same temperatures data set above but with many additional normal measurements:
\begin{gather*}
11,  11, 11, 11, 11, 11, 11, 11, 11, 11, 11, 11, 11, 11, 12, 12, 12, 12, 12, 12, 12, \\
 12,12, 12, 13, 13, 14, 14, 14, 17, 19, 19, 19, 19, 20, 21, 21, 21, 21, 21, 22, 23, \\
 24, 24, 24, 24, 26, 26, 30, 50, 55
\end{gather*}
The histogram and scatter plots are shown in Figure \ref{fig:temperatures2} including the same two anomalies as before. The IQR, z-score and perception all correctly identify the two anomalies $\{50, 55\}$. However, the modified z-score method flags too many normal measurements as anomalies:  $\{24, 24, 24, 24, 26, 26, 30, 50, 55\}$. This example was crafted to convey results of the modified z-score method that were observed for real-world data sets where the data may be heavily skewed. Note that even for this obvious non-Gaussian heavily skewed data the z-score and IQR methods still detect the two anomalies. 

\begin{figure}
\centering
  \includegraphics[width=1\linewidth]{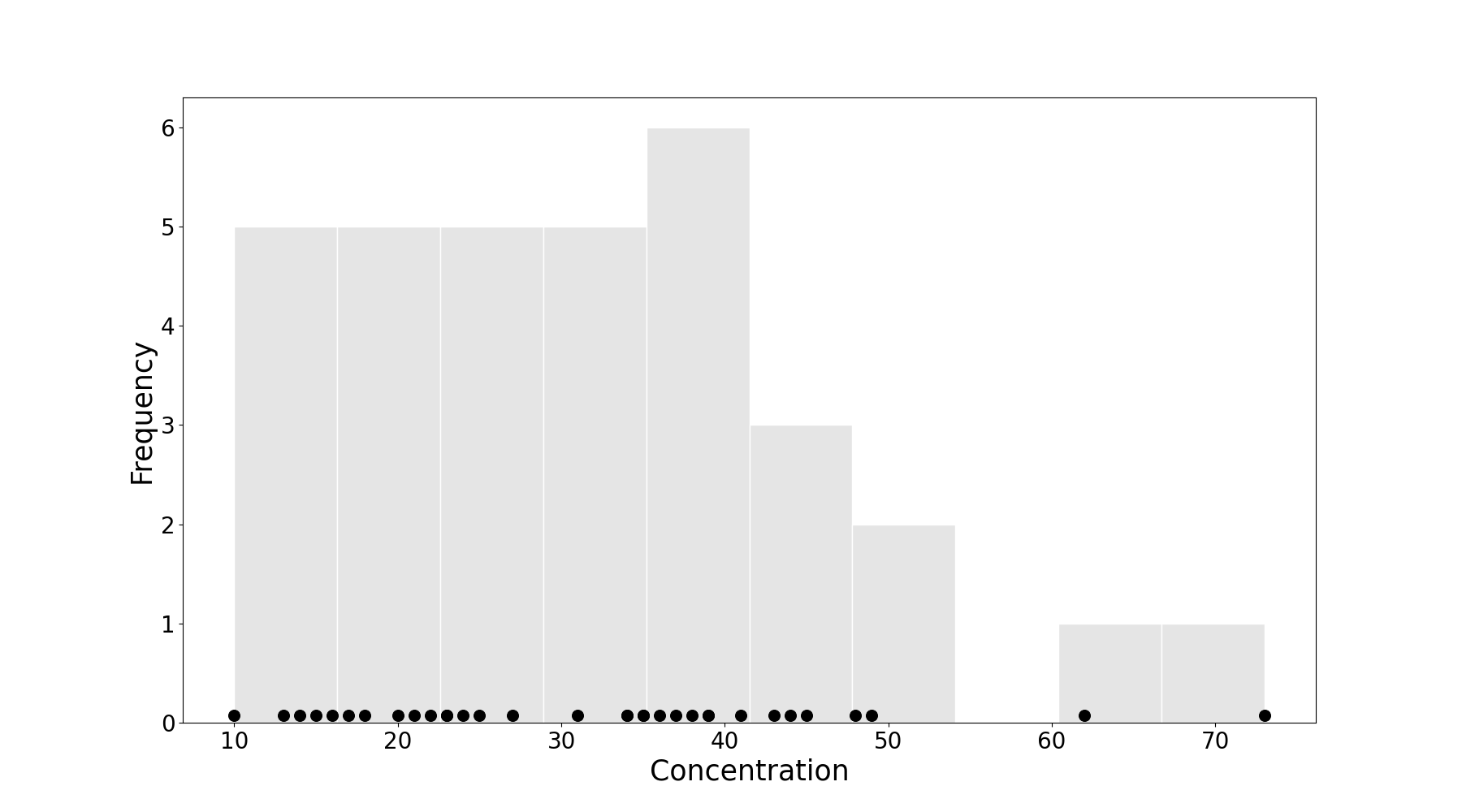}
  \caption{Distribution of measurements of lead concentration in the blood of children whose fathers worked in a battery manufacturing plant. This data set is taken from \cite{iglewicz1993detect} where there is an unusually high measurement of $\{73\}$ (and possibly also $\{62\}$). Anomalies detected: IQR, perception: $\{73\}$; modified z-score, z-score: none.}
  \label{fig:lead_exposure}
\end{figure}

The next example is another by \citet{iglewicz1993detect} where the authors provide measurements of the concentration of lead in the blood of children whose fathers worked in a battery manufacturing plant:
\begin{gather*}
10, 13, 14, 15, 16, 17, 18, 20, 21, 22, 23, 23, 24, 25, 27, 31, 34, 34, 35, 35, 36, \\
37, 38, 39, 39, 41, 43, 44, 45, 48, 49, 62, 73
\end{gather*}
The histogram and scatter plots are are shown in Figure \ref{fig:lead_exposure} where the perception and the IQR method flag the value $\{73\}$ as an anomaly. The value $\{62\}$ could also be considered a potential anomaly from looking at the scatter graph but its value is not found to be high enough. Interestingly both the z-score and MAD methods fail to detect any outliers in this data set.

\begin{figure}
\centering
  \includegraphics[width=1\linewidth]{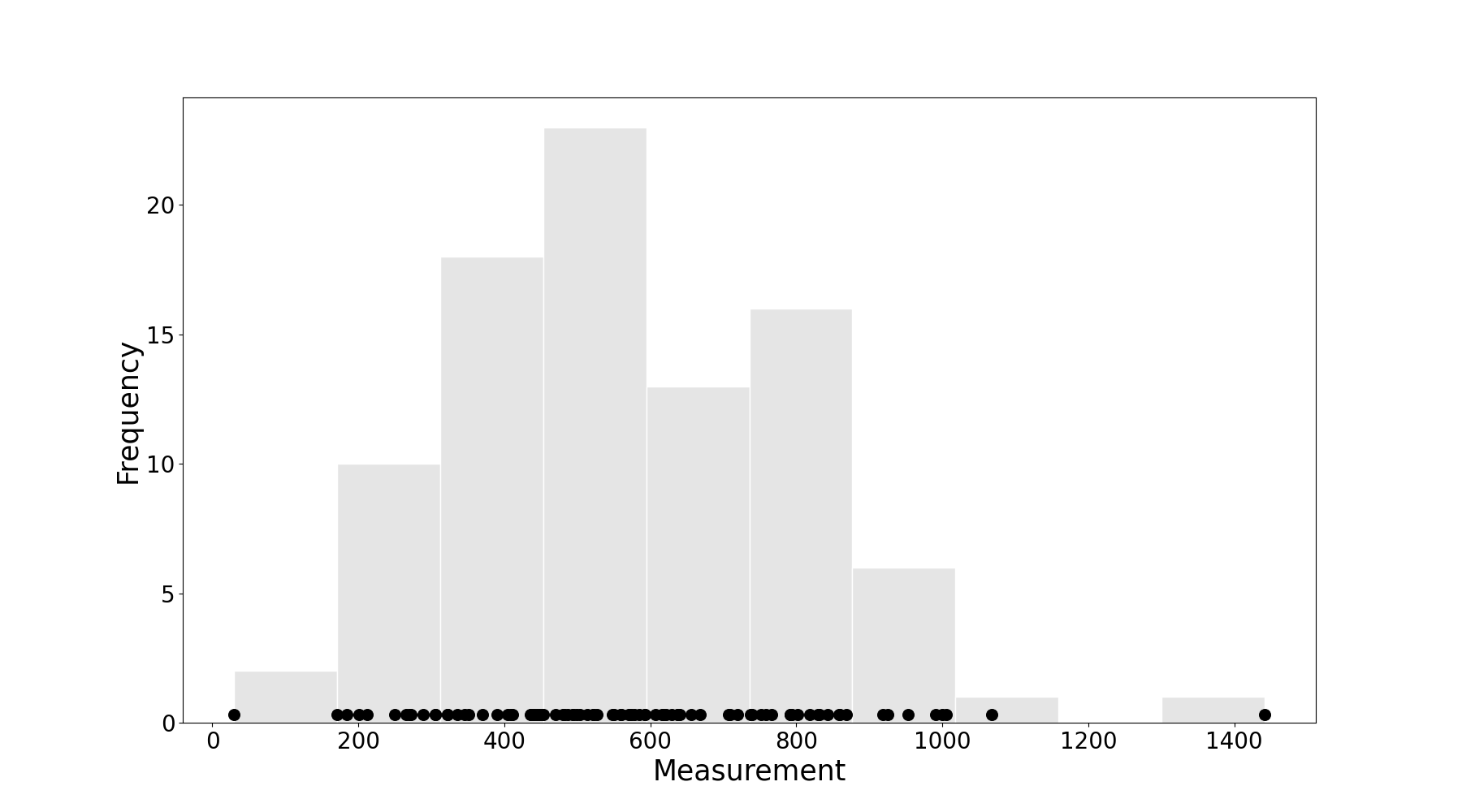}
  \caption{Distribution of data points to demonstrate outlier detection in \citet[chap. 7.1.6]{NISTeHandbook}. Anomalies detected: z-score, modified z-score, IQR: $\{1441\}$; perception: $\{30, 1068, 1441\}$.  
  }
  \label{fig:engineer_stats}
\end{figure}

\begin{figure}
\centering
  \includegraphics[width=1\linewidth]{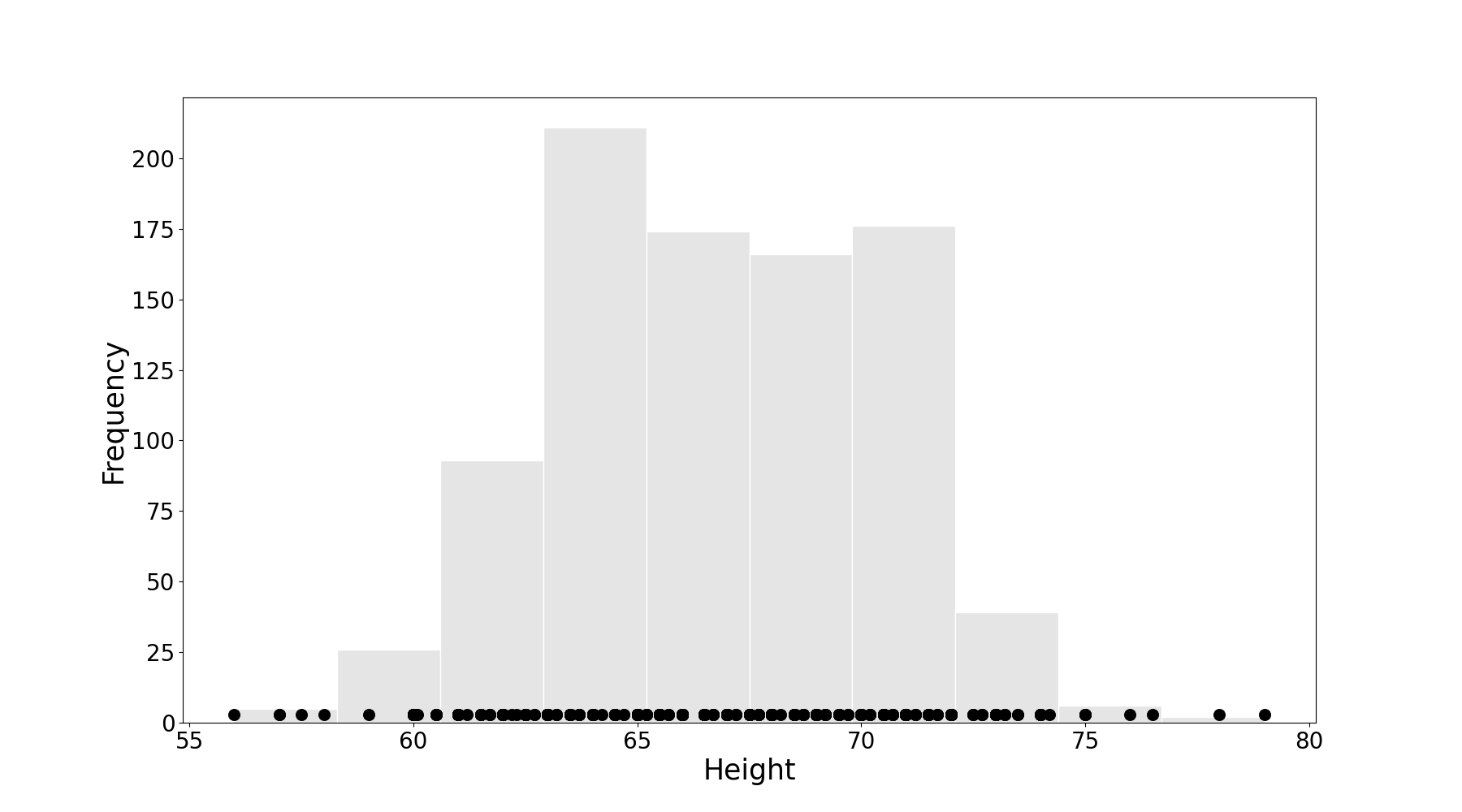}
  \caption{Galton height data set. Distribution of the classical height measurements recorded by Galton. Anomalies detected: modified z-score: none; IQR: $\{79\}$; z-score: $\{56, 78,79\}$; perception: $\{56, 57, 57.5, 76, 76.5, 78, 79\}$.  
}
  \label{fig:galton_heights}
\end{figure}

The final two examples are of medium sized data sets. The first is from \citet[chap. 7.1.6]{NISTeHandbook} where the histogram and scatter plots are shown in Figure \ref{fig:engineer_stats}. Two outliers can be perceived at values $\{30,1441\}$ which the perception algorithm detects, however, the z-score, MAD and IQR methods only find the value $\{1441\}$ to be an anomaly. Interestingly, given the data distribution, the perception algorithm also flags the value $\{1068\}$ as an anomaly which on closer inspection could be considered an outlier or a false positive. 
The second medium sized example is the classical Galton heights data set whose histogram and scatter plots are shown in Figure \ref{fig:galton_heights}. Given the large number of measurements and distribution of the data, a handful of outliers at either end of the tails of the distribution could be observed. The results here are interesting and surprising. The modified z-score method again does not flag any anomalies. The IQR method only flags the high end value of $\{79\}$ as an outlier. The z-score method only flags the outer values $\{56, 78,79\}$ to be anomalies. The perception algorithm selects more observations as anomalies at either end of the distribution tails: $\{56, 57, 57.5, 76, 76.5, 78, 79\}$. Which method here performs the best is subjective---are there in fact many, a few or no anomalies in the data? However, given just the data distribution by the histogram and scatter plot, I am inclined to say that the perception algorithm has selected more of the observations correctly as outliers that deserve further investigation since these are rarely occurring and found in the outer regions of the data.

The examples in this section highlight the varying results that can be obtained by different statistical anomaly detection algorithms on different data sets. The three traditionally used methods vary in their performance, but it can already be seen that the perception algorithm performs reliably over all the data sets to flag observations that could be deemed---at least visually---to most likely be anomalies. 

\subsubsection{Real World Large Data Sets}

This section presents results on large univariate data sets to compare the threshold boundaries of the anomaly detection algorithms. Due to the difficulties in obtaining publicly available univariate data sets I chose to use examples from the Numenta Anomaly Benchmark (NAB) \citep{NAB_Ahmad2017}. These are time-series data containing contextual and collective labelled anomalies. However, as the present work is only concerned with global point anomaly detection of particularly small or large values the labels are not used. 

\begin{figure}
\centering
  \includegraphics[width=\textwidth,height=\textheight,keepaspectratio]{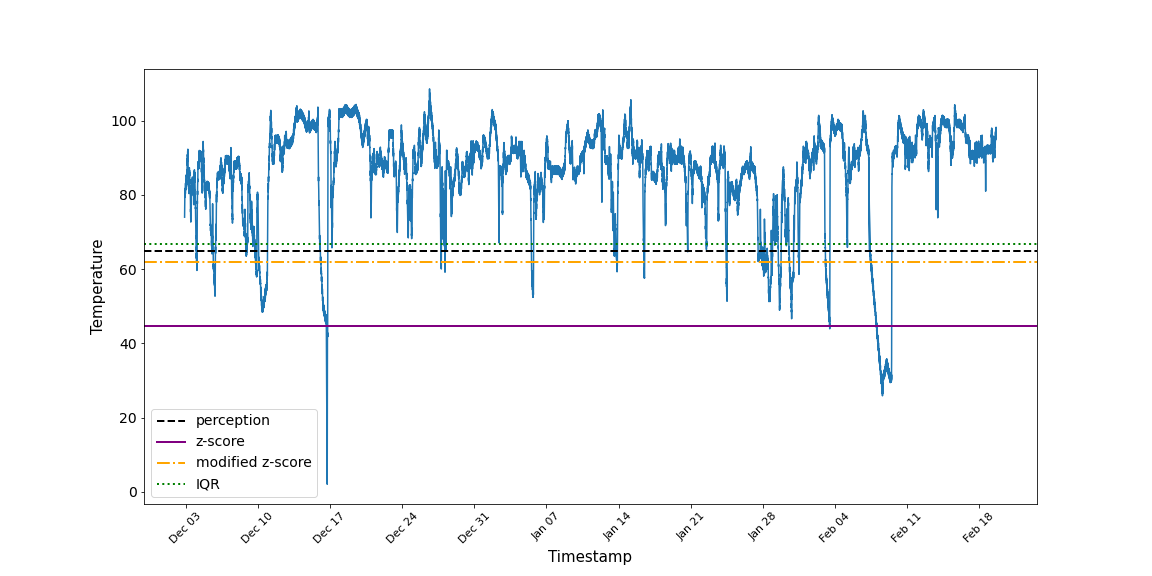}
  \caption{Temperature sensor data of an internal component of a large industrial machine taken over a period of a few months. Relatively low values (seen as sharp drops) are flagged as anomalies. Overlaid are the thresholds chosen by each method; below which all points are considered anomalies respectively.}
  \label{fig:machine_temperature}
\end{figure}

\begin{figure}
\centering
  \includegraphics [width=\textwidth,height=\textheight,keepaspectratio]{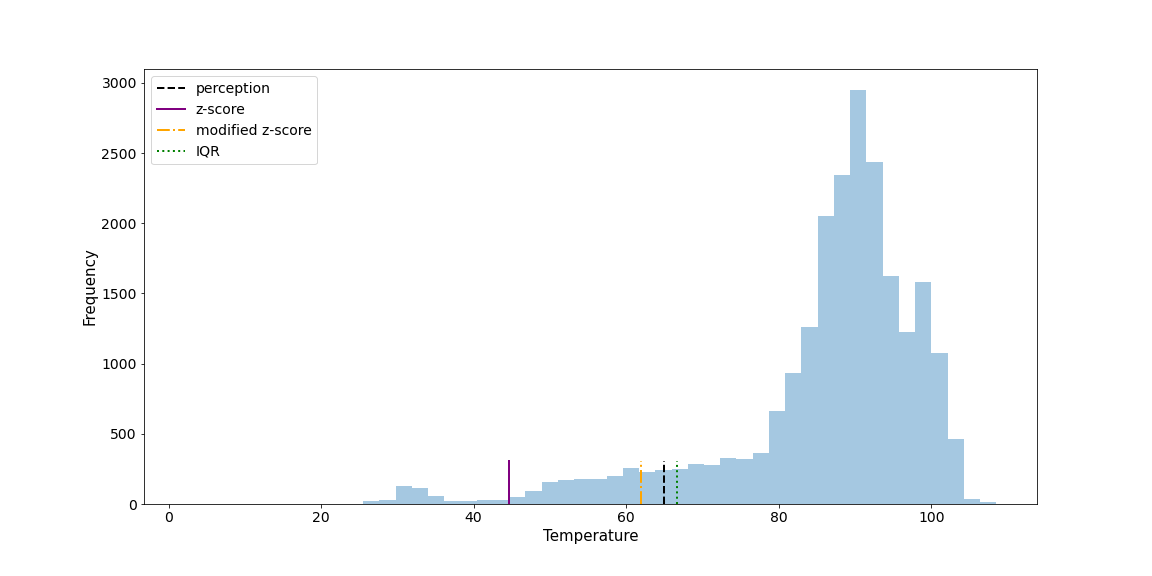}
  \caption{Temperature sensor data of an internal component of a large industrial machine taken over a period of a few months. The data is shown as a histogram and the threshold lines for each method are overlaid where all values to the left are considered anomalies repectively.}
  \label{fig:machine_temperature_dist}
\end{figure}

The first data set is of temperature sensor recordings from an internal component of a large industrial machine taken over a period of a few months. The line graph of the data is shown in Figure \ref{fig:machine_temperature} along with the threshold lines for each method; below which all points are flagged as anomalies. Figure \ref{fig:machine_temperature_dist} shows the same data but as a histogram with threshold lines, to the left of which points are considered anomalous respectively. Both figures show the perception, modified z-score and IQR all yield largely the same results to only flag relatively low valued observations as anomalous. However, the z-score method only detects more extreme anomalies which is expected given that outlying points heavily affect its decision boundaries.

\begin{figure}
\centering
  \includegraphics [width=\textwidth,height=\textheight,keepaspectratio]{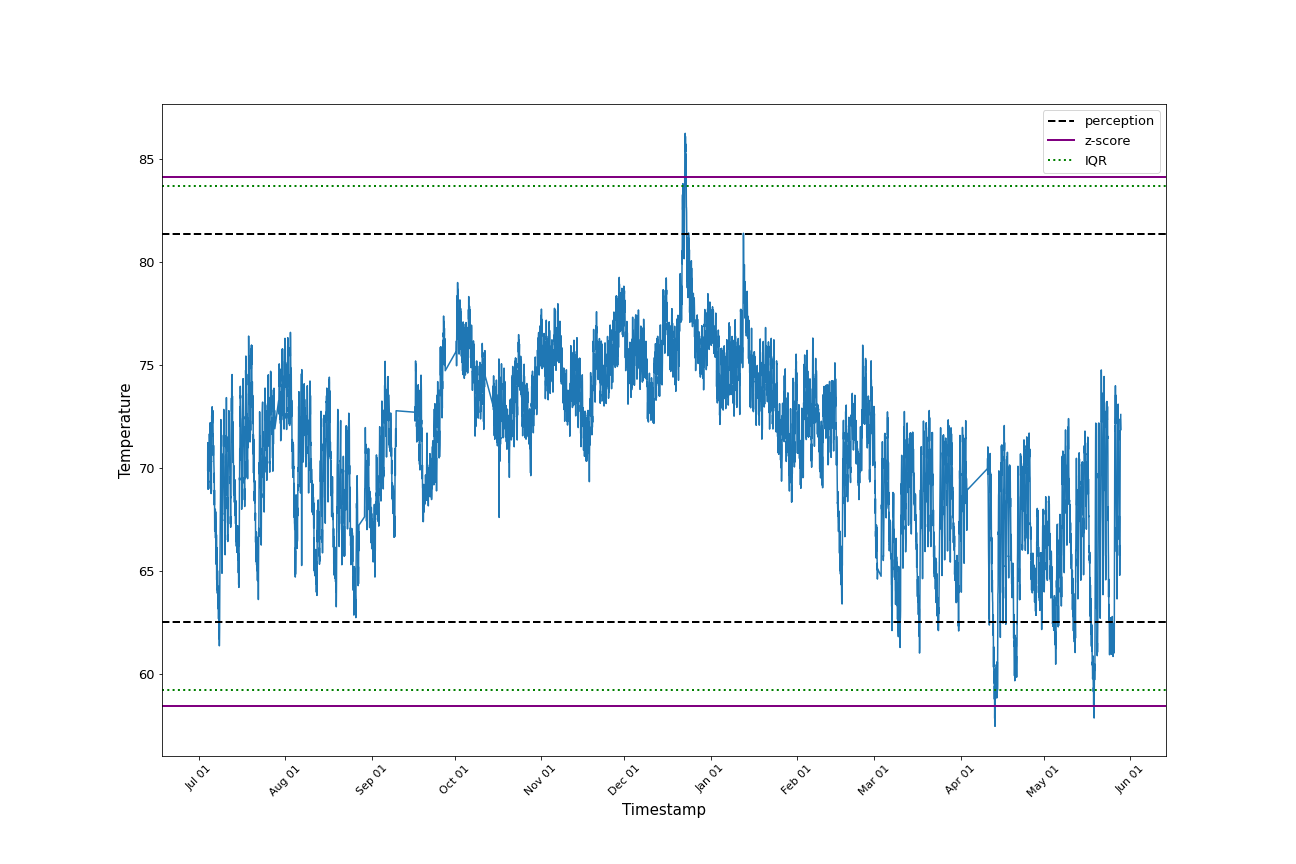}
  \caption{Ambient temperature measurements taken from an office environment over approximately a year. Unusually small or large values are flagged as anomalies in the data set. Overlaid on the graph are the thresholds chosen by each method where points \emph{above the top} and  \emph{below the bottom} set of lines are considered to be anomalies respectively.}
  \label{fig:ambient_temperature}
\end{figure}

\begin{figure}
\centering
  \includegraphics [width=\textwidth,height=\textheight,keepaspectratio]{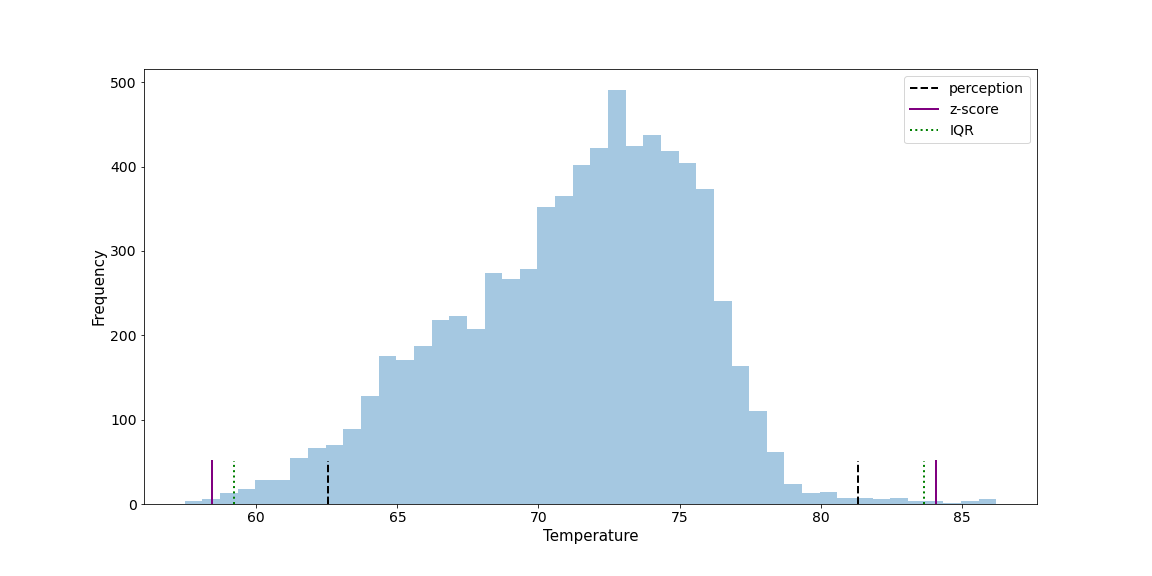}
  \caption{Ambient temperature measurements taken from an office environment over approximately a year. The data is shown as a histogram and the threshold lines at the tails of the distribution are overlaid for each method; beyond which temperature values are considered anomalous respectively.}
  \label{fig:ambient_temperature_dist}
\end{figure}

The second NAB data set is the ambient temperature measurements taken from an office environment over approximately a year. The line graph of the data is shown in Figure \ref{fig:ambient_temperature} along with the threshold lines for each method where any observation below the \emph{lower} line, and above the \emph{higher} line are flagged as anomalous by the respective method. Figure \ref{fig:ambient_temperature_dist} shows the same data but as a histogram where we observe the threshold lines again at the tails of the distribution. Visually we would be inclined to say that particularly low (or high) values in the data are anomalous with respect to the rest of the data points. The figures show that the perception algorithm flags more anomalies than the IQR and z-score methods, both of which flag only the more extreme points as anomalies. Interestingly, and as seen in the small univariate examples earlier, the modified z-score does not flag any points as anomalies even though there is a large spike in the line graph and infrequent points in the histogram. Although we have no ground truth, and the results here are only comparative, it is interesting to note that the modified z-score---often cited as being more robust and preferable over the z-score method---can completely fail to flag anomalies in some data sets. 

\begin{figure}
\centering
  \includegraphics[width=1\linewidth]{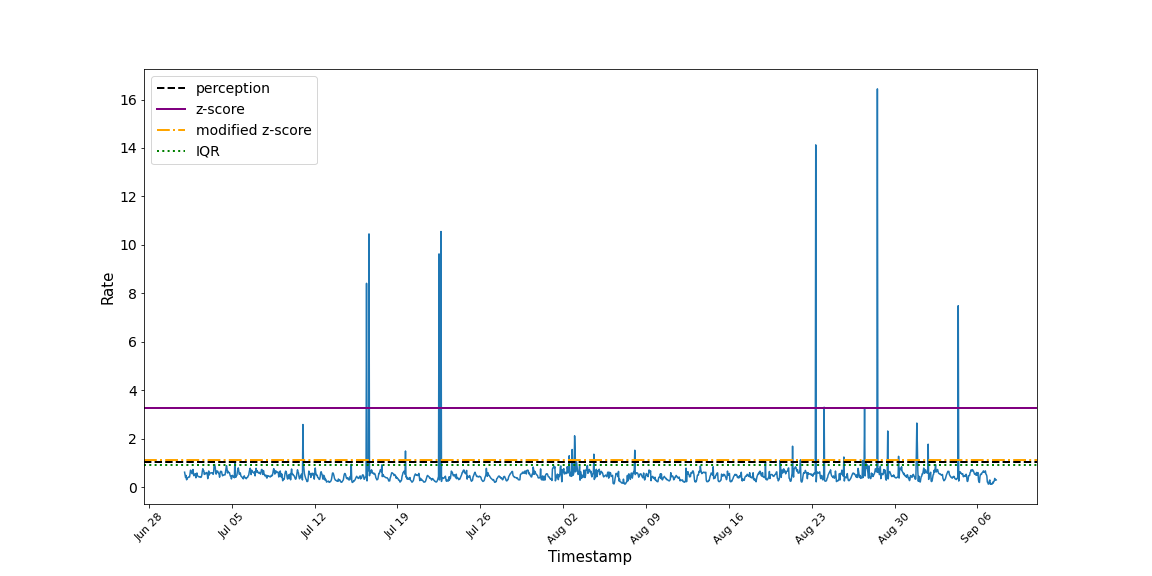}
  \caption{Online advertisement clicking rates are shown where there are relatively large spikes in the data that should be flagged as anomalies. Overlaid on the graph are the thresholds chosen by each method, above which all points are considered anomalies respectively.}
  \label{fig:adexchange_4}
\end{figure}

\begin{figure}
\centering
  \includegraphics[width=1\linewidth]{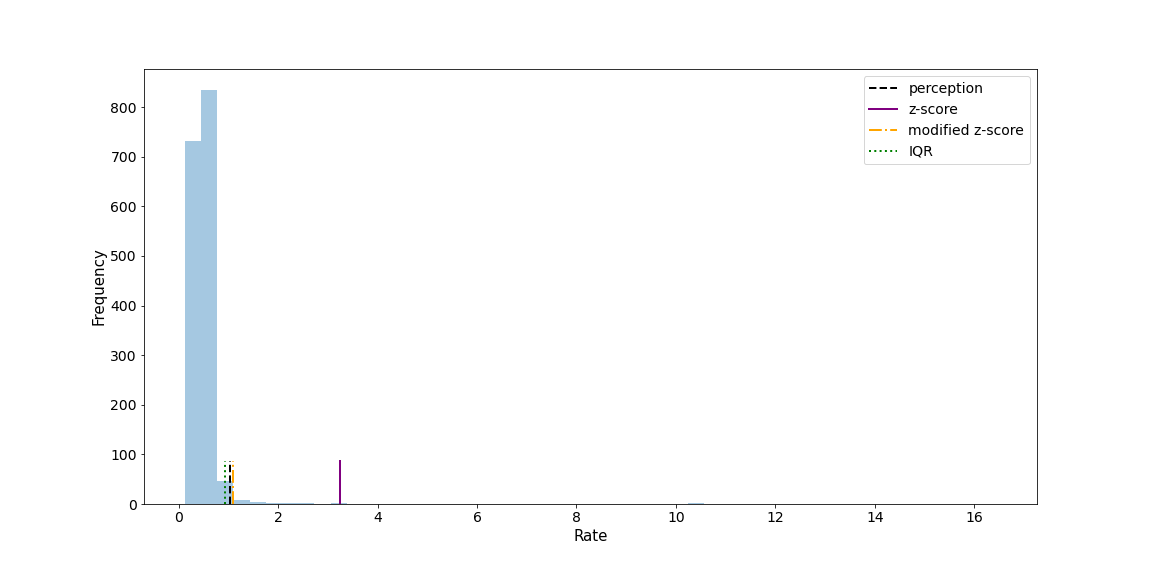}
  \caption{The histogram shows the distribution of online advertisement clicking rates with the threshold lines for each method overlaid; all values to the right are considered anomalous respectively.}
  \label{fig:adexchange_4_dist}
\end{figure}

The final example from the NAB data set is that of online advertisement clicking rates where the metric is cost per thousand impressions. Figure \ref{fig:adexchange_4} shows the line graph over time with the relatively large and rare peaks in the data. Overlaid are the threshold lines above which all points are considered anomalous with respect to that line. The modified z-score, IQR and perception all produce similar results. The z-score only considers much higher valued points to be anomalous and thus does not detect the smaller peaks even though they occur rarely. The histogram of the data is shown in Figure \ref{fig:adexchange_4_dist} where all points to the right of the overlaid threshold lines are considered anomalies respectively. The data is positively skewed and although the z-score selects points much further to the right of the main body of data, the other three methods place a similar and good threshold. 

The examples from NAB illustrate the results in detecting global point anomalies in large univariate real-world data. NAB has many more data sets where the perception algorithm was found to perform robustly on different distributions and ranges to report anomalies that match well with a visual inspection. With regards to the other methods, the z-score flags only more extreme anomalies than all other methods (as expected from section \ref{small_univariate_datasets}), the modified z-score gives good results but sometimes fails to detect what look like obvious anomalies in some examples (as was observed in Figures \ref{fig:lead_exposure} and \ref{fig:galton_heights}), while the IQR method performs well across the data sets although sometimes preferring to flag only more extreme points as anomalies. 

\subsection{Multivariate Data Results}
This section gives the results of the perception algorithm on multivariate data. The following assumptions (for the current version) limit the data sets selected and only need be approximately verified: 

\begin{itemize}
\item all feature values are numeric (other types such as nominal features are simply dropped),  
\item distribution of each feature value is unimodal where normal points largely form a single concentrated grouping, 
\item  anomalies compose $10\%$ or less of the data set in order to be rare,
\item anomalies are global i.e. they are to be found on the outlying regions of the distribution,
\item the representative features ensure anomalies are sufficiently different from the normal points to be separable.
\end{itemize}

Given such restrictions (it didn't limit the selection significantly), data sets can still greatly vary and it is recommended to test any new anomaly detection algorithm on a range of data sets to establish its average performance, and to also highlight its specific strengths and weaknesses. However, difficulties arise here with regards to finding labelled data sets due to their scarcity and their suitability to the task. Those that I chose were from \cite{Rayana:2016} and \cite{credit_card_fraud2015} which are modified from data sets originally used for gauging classification performance by supervised learning algorithms. In these data sets a set of classes are selected as the normal set and the rest considered to be anomalies or subsampled, or the data has been preprocessed for anonymization. An important remark is that these data sets are in some sense artificial due to the preprocessing decisions taken and hence are not organically representative of anomaly detection problems. Indeed, in private experiments it was found that for the same original data having been preprocessed by authors of different publications, the anomaly detection results can vary significantly.

The perception algorithm may be evaluated against a large number of multivariate anomaly detection algorithms. Hence I selected a variety of different approaches and only those that generally performed the best in studies and survey papers:

\begin{itemize}
\item Isolation Forest (IF)
\item One-class Support Vector Machine (OCSVM)
\item Minimum Covariance Determinant (MCD)
\item Local Outlier Factor (LOF) 
\item $K^{th}$-nearest neighbour (KNN) 
\item Histogram Based Outlier Selection (HBOS)
\end{itemize}

Focusing on unsupervised anomaly detection, the algorithms are only provided the data set with no labels or user acquired data specific hints for parameter choices. Thus any parameters that need to be set such as the ratio of outliers in the data set or the number of $k$ nearest neighbours must be chosen in advance. This is in line with real-world applications where we have little indication of parameter choices and of how many anomalies, if any, a given data set may contain; particularly in the case of streaming data. This causes problems for many learning algorithms since crucial parameters that significantly affect results are required for the user to specify. However, it merely highlights the difficulties of using such algorithms since not only is the implementation difficult, but so is the evaluation where many different parameters may have to be chosen and average results taken. Given the unsupervised nature of the tasks it was decided to keep to the default parameters of the pyOD \citep{zhao2019pyod} and Scikit-learn \citep{scikitlearn} anomaly detection libraries as a natural first choice.

The evaluation metrics selected are precision, recall, area under the ROC curve (AUC), F1-score and the runtime (total of initialisation, training and prediction times). The AUC score is the most cited metric for evaluating anomaly detection algorithms since it is a measure of how well the anomalies are ranked against the normal observations and has practical use for when an end user investigates the presented anomalies in order. However, in real-world scenarios, given a new data set, even if we can reasonably assume to have a high AUC score, selecting the appropriate threshold---akin to the ratio of anomalies---is difficult. Hence, it is perhaps not the best evaluation metric. It is a similar case for independently using precision, recall and the F1-score. Hence, any single metric used to state the best algorithm is avoided since the objective of success is dependent on application and because the criteria for success may not be captured by any one, or even all, of these metrics. 

\begin{figure}
\centering
  \includegraphics[width=1\linewidth]{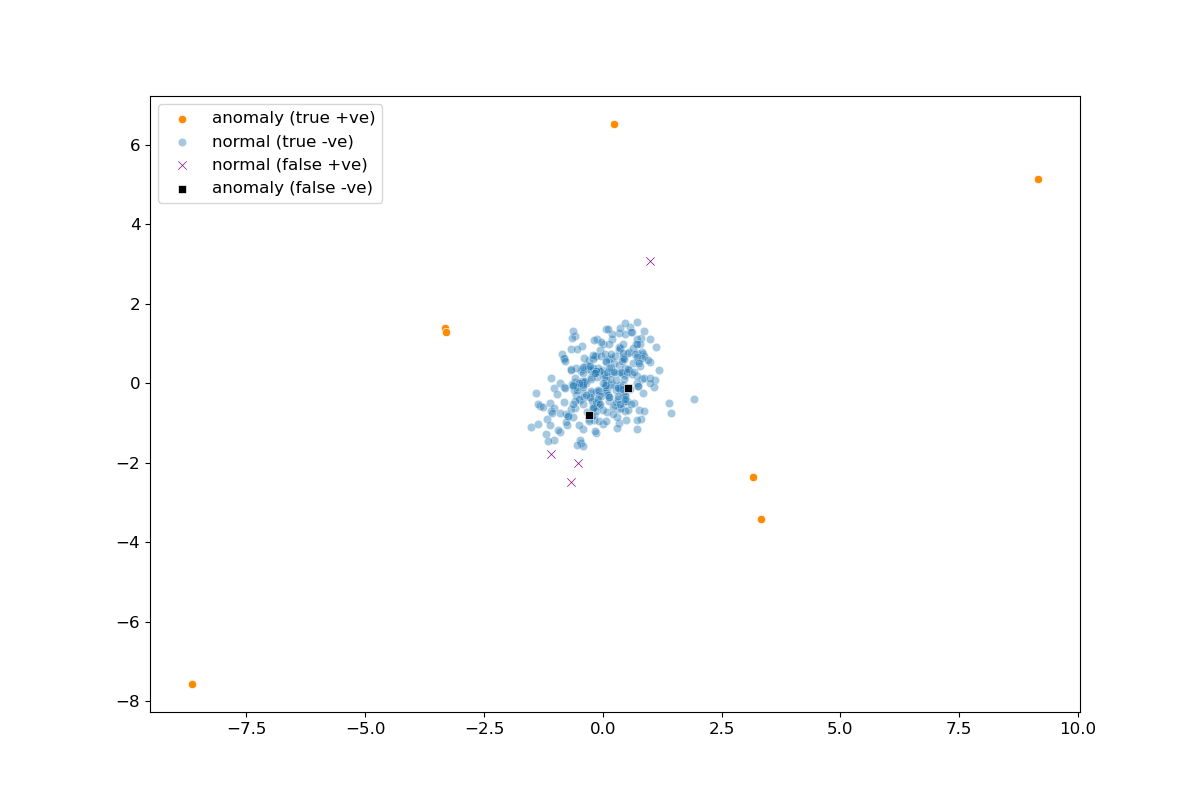}
  \caption{Perception algorithm results (validation data only). The behaviour of 307 servers measuring the throughput (mb/s) and latency (ms) of response from data set ex8data1 of \cite{NgML}. A central mass of points is observed with anomalies being generally distant from the cluster. Overlaid on the plot are the results of the perception algorithm. It correctly detects all the anomalies except for two that lie within the central cluster. A few false positives are also made due to the relative distance they are found from the cluster.}
  \label{fig:ex8data1_validation}
\end{figure}

\begin{figure}
\centering
  \includegraphics[width=1\linewidth]{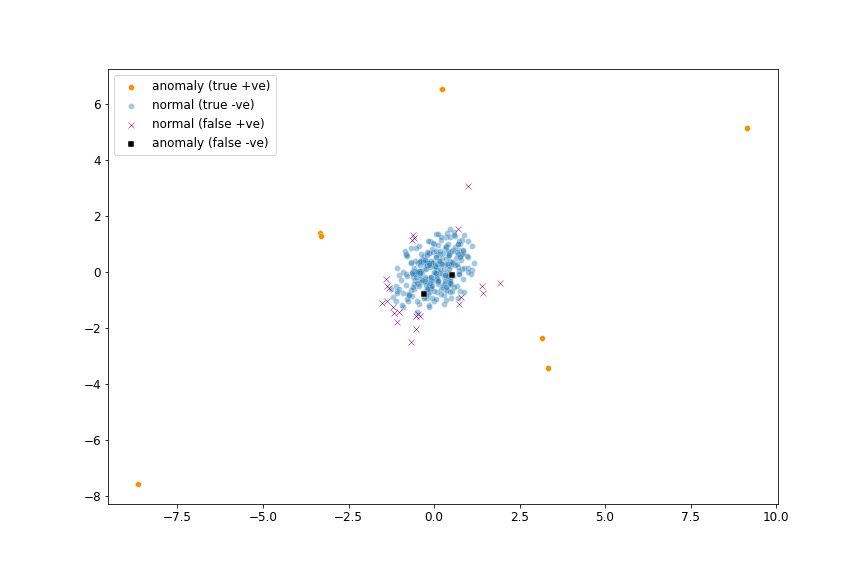}
  \caption{MCD model results (validation data only). The behaviour of 307 servers measuring the throughput (mb/s) and latency (ms) of response from data set ex8data1 of \cite{NgML}. Overlaid are the results of the MCD algorithm with default PyOD parameters where we clearly observe a high number of false positives as compared to the perception algorithm results in Figure \ref{fig:ex8data1_validation}.
  }
  \label{fig:ex8data1_validation_mcd}
\end{figure}

The first data set to consider is ex8data1 from \cite{NgML} which is a two dimensional sampling that measures the throughput (mb/s) and latency (ms) of response behaviours over 307 servers. The assumption here is that the majority of servers are operating normally with only a few that are behaving anomalously. The validation data set portion is used for both fitting the models and making predictions since there are plenty of data points and the full labels available where $9$ ($3\%$) of the observations are anomalies. Figure \ref{fig:ex8data1_validation} plots the standardised data set to illustrate the distribution of the data points where the results of the perception algorithm are overlaid. The normal measurements are located in a central cluster while the anomalies are mainly scattered further out. Most of the anomalies have been identified correctly except for the two (false negatives) shown within the central cluster that no algorithm can detect correctly given the data features. We also see a small number of false positives since the algorithm deems them to be sufficiently different from the normal points; these are borderline cases. For comparison, the results of the MCD method are displayed in Figure \ref{fig:ex8data1_validation_mcd}. The most striking difference is the large number of false positives produced; one reason is due to the default contamination ratio parameter of $0.1$ which specifies what percentage of points will be returned back as anomalies. The other algorithms produce similar results but select very different points as false positives. The numerical comparisons between all the algorithms are shown in Table \ref{table:ex8data1_validation_table} where we see the best precision, F1-score, AUC score and total runtime results are given by the perception algorithm while the recall scores are the same. This example provides a visual indicator of the performance of the perception algorithm in detecting anomalies without user specified parameters. 

\begin{table}
\robustify\bfseries
\centering
\begin{tabular}{llllll}
\toprule
Classifier &     Precision &        Recall &           F1 &           AUC &          Runtime \\
\midrule
      HBOS &          0.23 & \textbf{0.78} &         0.36 &          0.92 &          0.00141 \\
   IForest &          0.23 & \textbf{0.78} &         0.35 &          0.87 &          0.23463 \\
       KNN &          0.29 & \textbf{0.78} &         0.42 &          0.85 &          0.03347 \\
       LOF &          0.23 & \textbf{0.78} &         0.36 &          0.88 &          0.00421 \\
       MCD &          0.23 & \textbf{0.78} &         0.35 &          0.88 &          0.03292 \\
     OCSVM &          0.23 & \textbf{0.78} &         0.35 &          0.87 &          0.01994 \\
Perception & \textbf{0.64} & \textbf{0.78} & \textbf{0.7} & \textbf{0.93} & \textbf{0.00077} \\
\bottomrule
\end{tabular}

\caption{Results of unsupervised anomaly detection algorithms on the validation portion of data set ex8data1 from \cite{NgML}. Highlighted in bold is the best performance for a given metric. The perception algorithm performs the best here with respect to all metrics without requiring users to specify any parameters. The percentage of anomalies in this data set is $3\%$.}
\label{table:ex8data1_validation_table}
\end{table}

The next example is ex8data2 from \cite{NgML} that is composed of 11 features containing many more properties of computer servers and is claimed to be a more realistic data set to detect anomalies in than ex8data1. This data set consists of $1000$ unlabelled training examples that the models are fitted to and $100$ labelled examples that are predicted upon for evaluation purposes. The results are shown in Table \ref{table:ex8data2_validation_table} where there is no clear best performer across all the metrics. The perception algorithm has the best precision, F1-score and total runtime but with lowest recall and AUC score. It is important to note that the validation data set has $10\%$ anomalies which coincidentally corresponds to the default contamination ratio parameter that all the algorithms for comparison use. Thus, their results are optimal with respect to this parameter.

\begin{table}
\robustify\bfseries
\centering
\begin{tabular}{llllll}
\toprule
Classifier &     Precision &       Recall &            F1 &           AUC &          Runtime \\
\midrule
      HBOS &           0.5 &          0.7 &          0.58 &          0.93 &          0.00384 \\
   IForest &          0.64 &          0.7 & \textbf{0.67} &          0.93 &          0.29642 \\
       KNN &          0.44 &          0.7 &          0.54 &          0.93 &          0.03524 \\
       LOF &          0.44 &          0.7 &          0.54 &          0.93 &          0.02893 \\
       MCD &          0.44 & \textbf{0.8} &          0.57 &          0.94 &          0.53724 \\
     OCSVM &           0.5 & \textbf{0.8} &          0.62 & \textbf{0.95} &          0.09116 \\
Perception & \textbf{0.75} &          0.6 & \textbf{0.67} &          0.86 & \textbf{0.00165} \\
\bottomrule
\end{tabular}

  \caption{Results of unsupervised anomaly detection algorithms on data set ex8data2 from \cite{NgML}. Highlighted in bold is the best performance for a given metric. The perception algorithm performs the best here with respect to precision, F1-score and total runtime, but has the lowest recall and AUC scores. The percentage of anomalies in this test set is $10\%$.}
  \label{table:ex8data2_validation_table}
\end{table}

The final tables of results are for real-world anomaly detection data sets obtained from \cite{Rayana:2016} and \cite{credit_card_fraud2015}. Table \ref{table:data_set_properties} shows the selection along with their properties; the numbers of examples, features and percentages of anomalies vary significantly. Tables 4-8 show the results of applying the anomaly detection algorithms. These have been split by metrics and highlighted in bold are the top scores. The general findings are that the perception algorithm (implemented in python), consistently has the best precision and largely the best F1-score across the data sets by a significant margin. The recall scores are not significantly better, but comparable in some cases to the best performing algorithms, with no particular method standing out. The AUC scores are comparable with the best in the majority of data sets, but do require improvement in the rest. In terms of speed the perception algorithm is clearly very fast and competes with HBOS. Both algorithms show that they can produce results in real-time making them suitable for development on large data sets and streaming applications.

\begin{table}
\robustify\bfseries
\centering
\begin{tabular}{lrrr}
\toprule
       Name &  \# examples &  \# features &  \% anomalies \\
\midrule
credit-card &       284807 &           29 &          0.17 \\
     cardio &         1831 &           21 &          9.61 \\
    shuttle &        49097 &            9 &          7.15 \\
       musk &         3062 &          166 &          3.17 \\
       http &       567498 &            3 &          0.39 \\
       smtp &        95156 &            3 &          0.03 \\
    thyroid &         3772 &            6 &          2.47 \\
        wbc &          378 &           30 &          5.56 \\
 satimage-2 &         5803 &           36 &          1.22 \\
\bottomrule
\end{tabular}

  \caption{Properties of data sets selected from \cite{Rayana:2016} and \cite{credit_card_fraud2015} showing the varying numbers of examples, features and percentages of anomalies.}
  \label{table:data_set_properties}
\end{table}

\begin{table}
\robustify\bfseries
\centering
\resizebox{\textwidth}{!}{%
\begin{tabular}{llllllllll}
\toprule
Classifier &         cardio &    credit-card &           http &          musk &     satimage-2 &        shuttle &           smtp &        thyroid &            wbc \\
\midrule
      HBOS &          0.443 &          0.015 &          0.051 &         0.316 &          0.114 &          0.689 &          0.003 &          0.202 &          0.421 \\
   IForest &          0.568 &          0.015 &          0.039 &         0.316 &          0.119 &          0.704 &          0.002 &          0.233 &          0.395 \\
       KNN &           0.35 &            NaN &          0.002 &         0.139 &          0.093 &          0.208 &          0.003 &          0.238 &            0.4 \\
       LOF &           0.19 &            NaN &          0.001 &         0.137 &           0.04 &          0.115 &          0.003 &           0.06 &            0.4 \\
       MCD &          0.399 &          0.014 &          0.039 &         0.316 &          0.122 &          0.689 &          0.002 &          0.235 &          0.368 \\
     OCSVM &          0.503 &            NaN &            NaN &         0.316 &           0.12 &          0.697 &          0.002 &          0.214 &          0.395 \\
Perception & \textbf{0.591} & \textbf{0.031} & \textbf{0.149} & \textbf{0.89} & \textbf{0.333} & \textbf{0.907} & \textbf{0.006} & \textbf{0.247} & \textbf{0.483} \\
\bottomrule
\end{tabular}

}
  \caption{The precision results on the real-world data sets from \cite{Rayana:2016} and \cite{credit_card_fraud2015} are shown. $nan$ is used to indicate that the algorithm could not be run on the data within an acceptable time of 30 minutes. The perception algorithm achieves the best precision across all data sets usually by a significant margin.}
  \label{table:dataset_precision}
\end{table}

\begin{table}
\robustify\bfseries
\centering
\resizebox{\textwidth}{!}{%
\begin{tabular}{llllllllll}
\toprule
Classifier &        cardio &   credit-card &         http &         musk &   satimage-2 &       shuttle &          smtp &       thyroid &           wbc \\
\midrule
      HBOS &          0.46 & \textbf{0.89} & \textbf{1.0} & \textbf{1.0} &         0.93 &          0.96 &           0.7 &          0.82 & \textbf{0.76} \\
   IForest & \textbf{0.59} & \textbf{0.89} & \textbf{1.0} & \textbf{1.0} &         0.97 & \textbf{0.98} & \textbf{0.77} &          0.95 &          0.71 \\
       KNN &          0.28 &           NaN &         0.04 &         0.27 &         0.66 &          0.22 &          0.73 &          0.87 &          0.67 \\
       LOF &          0.16 &           NaN &         0.02 &         0.38 &         0.28 &          0.14 &           0.7 &          0.22 &          0.67 \\
       MCD &          0.41 &          0.82 & \textbf{1.0} & \textbf{1.0} & \textbf{1.0} &          0.96 & \textbf{0.77} & \textbf{0.96} &          0.67 \\
     OCSVM &          0.52 &           NaN &          NaN & \textbf{1.0} &         0.99 &          0.97 & \textbf{0.77} &          0.87 &          0.71 \\
Perception &          0.37 &          0.88 & \textbf{1.0} & \textbf{1.0} &          0.9 &          0.96 &           0.7 &          0.68 &          0.67 \\
\bottomrule
\end{tabular}

}
  \caption{The recall results on the data sets from \cite{Rayana:2016} and \cite{credit_card_fraud2015} are shown. $nan$ is used to indicate that the algorithm could not be run on the data within an acceptable time of 30 minutes. The perception algorithm generally achieves recall amongst the top scores, but not always the best. While there isn't a clear best performer, the algorithms have similar results where they could be applied, except for LOF and KNN which have poorer results.}
  \label{table:dataset_recall}
\end{table}

\begin{table}
\robustify\bfseries
\centering
\resizebox{\textwidth}{!}{%
\begin{tabular}{llllllllll}
\toprule
Classifier &         cardio &    credit-card &          http &           musk &     satimage-2 &        shuttle &           smtp &        thyroid &           wbc \\
\midrule
      HBOS &          0.451 &           0.03 &         0.097 &           0.48 &          0.202 &          0.801 &          0.005 &          0.323 &         0.542 \\
   IForest & \textbf{0.579} &           0.03 &         0.075 &           0.48 &          0.212 &          0.821 &          0.005 &          0.374 &         0.508 \\
       KNN &           0.31 &            NaN &         0.004 &          0.183 &          0.163 &          0.214 &          0.005 &          0.374 &           0.5 \\
       LOF &          0.176 &            NaN &         0.002 &          0.201 &           0.07 &          0.126 &          0.005 &          0.094 &           0.5 \\
       MCD &          0.407 &          0.028 &         0.075 &           0.48 &          0.218 &          0.803 &          0.005 & \textbf{0.378} &         0.475 \\
     OCSVM &          0.513 &            NaN &           NaN &           0.48 &          0.215 &          0.812 &          0.005 &          0.344 &         0.508 \\
Perception &          0.455 & \textbf{0.061} & \textbf{0.26} & \textbf{0.942} & \textbf{0.487} & \textbf{0.931} & \textbf{0.012} &          0.362 & \textbf{0.56} \\
\bottomrule
\end{tabular}

}
  \caption{The F1-score results on the data sets from \cite{Rayana:2016} and \cite{credit_card_fraud2015} are shown. $nan$ is used to indicate that the algorithm could not be run on the data within an acceptable time of 30 minutes. The perception algorithm generally has the best F1-score; sometimes significantly so, other times on a similar level.}
  \label{table:dataset_f1-score}
\end{table}

\begin{table}
\robustify\bfseries
\centering
\resizebox{\textwidth}{!}{%
\begin{tabular}{llllllllll}
\toprule
Classifier &        cardio &   credit-card &         http &         musk &   satimage-2 &      shuttle &          smtp &       thyroid &           wbc \\
\midrule
      HBOS &          0.85 & \textbf{0.95} &         0.99 & \textbf{1.0} &         0.98 &         0.98 &           0.8 &          0.95 & \textbf{0.96} \\
   IForest & \textbf{0.94} & \textbf{0.95} & \textbf{1.0} & \textbf{1.0} &         0.99 & \textbf{1.0} &          0.91 &          0.98 & \textbf{0.96} \\
       KNN &          0.69 &           NaN &         0.25 &         0.62 &         0.93 &         0.63 &          0.91 &          0.96 &          0.95 \\
       LOF &          0.55 &           NaN &          0.4 &         0.64 &         0.54 &         0.52 &          0.83 &          0.66 &          0.94 \\
       MCD &           0.8 &          0.92 & \textbf{1.0} & \textbf{1.0} & \textbf{1.0} &         0.99 & \textbf{0.95} & \textbf{0.99} &          0.92 \\
     OCSVM & \textbf{0.94} &           NaN &          NaN & \textbf{1.0} & \textbf{1.0} &         0.99 &          0.85 &          0.96 &          0.94 \\
Perception &          0.77 &          0.93 & \textbf{1.0} & \textbf{1.0} &         0.93 &         0.98 &           0.8 &          0.86 &          0.76 \\
\bottomrule
\end{tabular}

}
 \caption{The AUC-score results on the data sets from \cite{Rayana:2016} and \cite{credit_card_fraud2015} are shown. $nan$ is used to indicate that the algorithm could not be run on the data within an acceptable time of 30 minutes. Of the best best performing algorithms, the scores are quite similar with no particular algorithm standing out. The perception algorithm performs comparably in the majority of data sets.}
  \label{table:dataset_auc}
\end{table}

\begin{table}
\robustify\bfseries
\centering
\resizebox{\textwidth}{!}{%
\begin{tabular}{llllllllll}
\toprule
Classifier &         cardio &    credit-card &           http &           musk &     satimage-2 &        shuttle &           smtp &        thyroid &            wbc \\
\midrule
      HBOS &          0.007 &          0.557 & \textbf{0.167} &          0.073 &          0.094 & \textbf{0.022} & \textbf{0.035} & \textbf{0.004} &          0.013 \\
   IForest &          0.359 &         33.445 &         46.118 &          0.939 &          1.121 &          3.695 &          7.788 &          0.444 &          0.401 \\
       KNN &          0.347 &            NaN &          73.41 &          3.119 &          2.408 &         17.525 &         10.226 &          0.495 &          0.055 \\
       LOF &          0.131 &            NaN &         25.678 &          0.453 &          1.443 &         23.157 &          2.345 &          0.218 &           0.01 \\
       MCD &          0.486 &         33.926 &         44.186 &          9.901 &          2.928 &         14.148 &         20.713 &          1.451 &          0.056 \\
     OCSVM &          0.561 &            NaN &            NaN &          2.362 &          7.586 &        384.331 &       1464.868 &          2.141 &          0.031 \\
Perception & \textbf{0.002} & \textbf{0.394} &          0.537 & \textbf{0.013} & \textbf{0.009} &          0.053 &          0.086 &          0.005 & \textbf{0.001} \\
\bottomrule
\end{tabular}

}
  \caption{The total initialisation, training and prediction runtime in seconds by the algorithms on the data sets from \cite{Rayana:2016} and \cite{credit_card_fraud2015} are shown. $nan$ is used to indicate that the algorithm could not be run on the data within an acceptable time of 30 minutes. Both HBOS and the perception algorithm complete exceptionally fast which illustrate their applicability to real-time anomaly detection.} 
  \label{table:dataset_total_time}
\end{table}

\section{Conclusion and Future Work}
\label{section:conclusions} 
%

This paper makes two primary contributions to the field of anomaly or outlier detection. The first is a definition of what an anomaly is and what should be computed for their detection. This is inspired by the brilliant capability of human vision to spot anomalies, and developed from principles of human perception---mainly the Helmholtz principle---that vision innately perceives certain a priori groupings as meaningful if their expectation of occurrence is low in uniform random noise. These groupings are taken to be the \emph{gestalt groupings} from the Gestalt Theory of Psychology and include colour, shape, proximity, good continuity, prior experience, closure and symmetry. The fundamental premise is that anomalies are those observations that unexpectedly vary by a particular feature of interest with respect to a main grouping of observations. Indeed, when a suitable measure is composed, it is shown that where the expectation of occurrence of a random variable is $<1$, a perception occurs to view the event as meaningful or in other words, an anomaly. 

The second primary contribution is the perception algorithm which directly finds anomalies in numerical multivariate data by a simple, fast yet highly effective method without modelling the normal data. The algorithm makes no assumption about the exact data distribution except (in the current version) that the grouping of normal observations concentrates in a single body and that anomalies are global and rare. The central idea is to assume observations unexpectedly distant from a point of centrality are anomalies, and to further assume that such distance values are generated by a stream of binary indicators with each value representing a summation of indicators from the stream in adjacent windows of equal length. Under an assumption that the non-zero indicators are uniformly, randomly and independently distributed amongst the windows, an unexpectedly large number appearing in a window is deemed to be an outlier event with respect to the rest of the observations. Crucially the algorithm is parameter-free by the natural use of expectations instead of probabilities where if the event occurs, but its expectation of occurrence is $<1$, then it is deemed unexpected, and hence an anomaly. This results in a \emph{truly} unsupervised learning algorithm since users only need supply data in the appropriate format and not additional data specific parameters. 

Results on univariate and multivariate data are promising. Indeed, for univariate data the visual graphs and algorithm output indicates that there is a case for using the perception algorithm in addition to existing methods used in the sciences, such as the z-score, modified z-score and IQR method. Another supporting factor is that the perception algorithm is applied parameter-free and hence not subject to careless or unscrupulous practice where users select parameters that give the desired outcome in experiments. However, such statements will require theoretical support in addition to the empirical findings. In the case of multivariate data the evaluation is more challenging due to the variety and dimensionality of data features, and the unavailability of true anomaly detection data sets for benchmarking algorithms. However, results show that the perception algorithm performs admirably well by current metrics. In general the perception algorithm showed the best precision, F1-score and had good recall and AUC scores---keeping in mind that the competing algorithms were left at default library parameters. The algorithm is also lightweight and runs exceptionally quick making it suitable for resource constrained and real-time applications.        

The philosophical approach based on principles of human perception, together with the experimental results, supports an alternative perspective for the detection of anomalies over traditional statistical and computational heuristic based methods. The general approach is mathematically derived from first principles that does not assume a typical (and not always true) Gaussian distribution of the data, and the use of expectations instead of probabilities leads to an elegant solution, while the practical aspects of the perception algorithm mean its application is straightforward. The results on publicly available data sets are good in comparison to existing solutions; however there is clearly room for improvement across all evaluation metrics and several issues to consider. Firstly, the algorithm is yet to be fully evaluated in real world production environments such as cyber intrusion detection systems which will give an additional layer of validation. Secondly, while specific modelling of the data is not carried out by the algorithm, representation of the data in an appropriate format is still required from the user. For example, vector valued points may not always be the most appropriate way to represent data. Indeed, in some problems like finding outliers in regression data it may be more appropriate to model distances of points from a plane of best fit to uncover outliers of interest. Thirdly, the normal data needs to form one central grouping in order to detect global anomalies, hence it can fail where there are clusters of normal points and anomalies lie in between. Fourthly, extremely large anomalies can affect the algorithm decision boundary leading to the masking effect where anomalies that would otherwise have been detected are instead considered to be normal points. These issues will be addressed in subsequent work together with developing methods for multivariate anomaly detection that improve upon the current results---particularly in the case of big data and high dimensional data sets. 


\section{Acknowledgements}
I would like to thank Felix Meier-Hedde for valuable insight and conversations regarding anomaly detection for cyber security applications. This research was funded by Endeavr Wales and Airbus.


\bibliography{Bibliography} 

\begin{thebibliography}{27}
\providecommand{\natexlab}[1]{#1}
\providecommand{\url}[1]{\texttt{#1}}
\expandafter\ifx\csname urlstyle\endcsname\relax
  \providecommand{\doi}[1]{doi: #1}\else
  \providecommand{\doi}{doi: \begingroup \urlstyle{rm}\Url}\fi

\bibitem[Aggarwal(2017)]{AggarwalOutlierAnalysis2017}
Charu~C. Aggarwal.
\newblock \emph{Outlier Analysis}.
\newblock Springer Publishing Company, Incorporated, 2nd edition, 2017.
\newblock ISBN 3319475770.

\bibitem[Aggarwal and Sathe(2017)]{AggarawalEnsembles2017}
Charu~C. Aggarwal and Saket Sathe.
\newblock \emph{Outlier Ensembles: An Introduction}.
\newblock Springer Publishing Company, Incorporated, 1st edition, 2017.
\newblock ISBN 331954764X.

\bibitem[Ahmad et~al.(2017)Ahmad, Lavin, Purdy, and Agha]{NAB_Ahmad2017}
Subutai Ahmad, Alexander Lavin, Scott Purdy, and Zuha Agha.
\newblock Unsupervised real-time anomaly detection for streaming data.
\newblock \emph{Neurocomputing}, 2017.

\bibitem[Barnett and Lewis(1978)]{Barnett1978}
Vic Barnett and Toby Lewis.
\newblock \emph{Outliers in statistical data.}
\newblock John Wiley \& Sons Ltd., 2nd edition, 1978.

\bibitem[Breunig et~al.(2000)Breunig, Kriegel, Ng, and Sander]{Breunig2000}
Markus Breunig, Hans-Peter Kriegel, Raymond~T. Ng, and J\"{o}rg Sander.
\newblock Lof: Identifying density-based local outliers.
\newblock \emph{SIGMOD Rec.}, 29\penalty0 (2):\penalty0 93--104, May 2000.

\bibitem[Chandola et~al.(2009)Chandola, Banerjee, and Kumar]{Chandola2009}
Varun Chandola, Arindam Banerjee, and Vipin Kumar.
\newblock Anomaly detection: A survey.
\newblock \emph{ACM Comput. Surv.}, 41\penalty0 (3), July 2009.

\bibitem[Desolneux et~al.(2007)Desolneux, Moisan, and Morel]{Morel07}
Agnès Desolneux, Lionel Moisan, and Jean-Michel Morel.
\newblock \emph{From Gestalt Theory to Image Analysis: A Probabilistic
  Approach}.
\newblock Springer Publishing Company, Incorporated, 1st edition, 2007.
\newblock ISBN 0387726357.

\bibitem[Ester et~al.(1996)Ester, Kriegel, Sander, and Xu]{Ester96}
Martin Ester, Hans-Peter Kriegel, J\"{o}rg Sander, and Xiaowei Xu.
\newblock A density-based algorithm for discovering clusters in large spatial
  databases with noise.
\newblock In \emph{Proceedings of the Second International Conference on
  Knowledge Discovery and Data Mining}, KDD'96, page 226–231. AAAI Press,
  1996.

\bibitem[Goldstein and Dengel(2012)]{goldstein2012}
Markus Goldstein and Andreas Dengel.
\newblock Histogram-based outlier score (hbos): A fast unsupervised anomaly
  detection algorithm.
\newblock 2012.

\bibitem[Goldstein and Uchida(2016)]{Goldstein2016}
Markus Goldstein and Seiichi Uchida.
\newblock A comparative evaluation of unsupervised anomaly detection algorithms
  for multivariate data.
\newblock \emph{PLOS ONE}, 11:\penalty0 1--31, 04 2016.

\bibitem[Hawkins(1980)]{Hawkins80}
Douglas~M. Hawkins.
\newblock \emph{Identification of Outliers}.
\newblock Chapman and Hall London; New York, 1980.

\bibitem[Iglewicz and Hoaglin(1993)]{iglewicz1993detect}
Boris Iglewicz and David~C. Hoaglin.
\newblock \emph{How to Detect and Handle Outliers}.
\newblock ASQC basic references in quality control. ASQC Quality Press, 1993.

\bibitem[Liu et~al.(2008)Liu, Ting, and Zhou]{Liu2008}
Fei~Tony Liu, Kai~Ming Ting, and Zhi-Hua Zhou.
\newblock Isolation forest.
\newblock In \emph{Proceedings of the 2008 Eighth IEEE International Conference
  on Data Mining}, ICDM, pages 413--422, Washington, DC, USA, 2008. IEEE
  Computer Society.

\bibitem[Lowe(1985)]{Lowe85}
David Lowe.
\newblock Perceptual organization and visual recognition, 1985.

\bibitem[MacQueen(1967)]{macqueen1967}
James MacQueen.
\newblock Some methods for classification and analysis of multivariate
  observations.
\newblock In \emph{Proc. of the fifth Berkeley Symposium on Mathematical
  Statistics and Probability}, volume~1, pages 281--297. University of
  California Press, 1967.

\bibitem[Marr(1982)]{Mar82}
David Marr.
\newblock \emph{Vision: A Computational Investigation into the Human
  Representation and Processing of Visual Information}.
\newblock Henry Holt and Co., Inc., USA, 1982.

\bibitem[Metzger(1975)]{Metz75}
Wolfgang Metzger.
\newblock \emph{Gesetze des Sehens}.
\newblock Waldemar Kramer, 1975.

\bibitem[Ng(2012)]{NgML}
Andrew Ng.
\newblock \emph{Machine learning: Programming Exercise 8: Anomaly Detection and
  Recommender Systems}.
\newblock 2012.

\bibitem[NIST(2012)]{NISTeHandbook}
NIST.
\newblock Engineering statistics handbook.
\newblock Technical report, U.S. Department of Commerce, 2012.
\newblock URL \url{https://books.google.co.uk/books?id=v-XXjwEACAAJ}.

\bibitem[Pedregosa et~al.(2011)Pedregosa, Varoquaux, Gramfort, Michel, Thirion,
  Grisel, Blondel, Prettenhofer, Weiss, Dubourg, Vanderplas, Passos,
  Cournapeau, Brucher, Perrot, and Duchesnay]{scikitlearn}
Fabian Pedregosa, Gael Varoquaux, Alexandre Gramfort, Vincent Michel, Bertrand
  Thirion, Olivier Grisel, Mathieu Blondel, Peter Prettenhofer, Ron Weiss,
  Vincent Dubourg, Jake Vanderplas, Alexandre Passos, David Cournapeau,
  Matthieu Brucher, Matthieu Perrot, and Edouard Duchesnay.
\newblock Scikit-learn: Machine learning in python.
\newblock \emph{Journal of Machine Learning Research}, 12:\penalty0 2825--2830,
  2011.

\bibitem[Pozzolo et~al.(2015)Pozzolo, Caelen, Johnson, and
  Bontempi]{credit_card_fraud2015}
Andrea~Dal Pozzolo, Olivier Caelen, Reid~A. Johnson, and Gianluca Bontempi.
\newblock Calibrating probability with undersampling for unbalanced
  classification.
\newblock In \emph{2015 IEEE Symposium Series on Computational Intelligence},
  pages 159--166, 2015.

\bibitem[Ramaswamy et~al.(2000)Ramaswamy, Rastogi, and Shim]{Ramaswamy2000}
Sridhar Ramaswamy, Rajeev Rastogi, and Kyuseok Shim.
\newblock Efficient algorithms for mining outliers from large data sets.
\newblock \emph{SIGMOD Rec.}, 29\penalty0 (2), May 2000.

\bibitem[Rayana(2016)]{Rayana:2016}
Shebuti Rayana.
\newblock Odds library.
\newblock Stony Brook University, Department of Computer Sciences, 2016.
\newblock URL \url{http://odds.cs.stonybrook.edu}.

\bibitem[Rousseeuw and Driessen(1999)]{mcd99}
Peter Rousseeuw and Katrien Driessen.
\newblock A fast algorithm for the minimum covariance determinant estimator.
\newblock \emph{Technometrics}, 41:\penalty0 212--223, 08 1999.

\bibitem[Schölkopf et~al.(2001)Schölkopf, Platt, Shawe-Taylor, Smola, and
  Williamson]{sch99}
Bernhard Schölkopf, John Platt, John Shawe-Taylor, Alexander Smola, and Robert
  Williamson.
\newblock Estimating support of a high-dimensional distribution.
\newblock \emph{Neural Computation}, 13:\penalty0 1443--1471, 07 2001.

\bibitem[Wertheimer(1923)]{Wer23}
Max Wertheimer.
\newblock \emph{Untersuchungen zur lehre der gestalt}, volume~4.
\newblock Psychologische Forschung, 1923.

\bibitem[Zhao et~al.(2019)Zhao, Nasrullah, and Li]{zhao2019pyod}
Yue Zhao, Zain Nasrullah, and Zheng Li.
\newblock Pyod: A python toolbox for scalable outlier detection.
\newblock \emph{Journal of Machine Learning Research}, 20\penalty0
  (96):\penalty0 1--7, 2019.

\end{thebibliography}

\end{document}